\documentclass[twoside, 12pt]{article}

\usepackage{graphicx, subfigure}
\usepackage[top=1in,bottom=1in,left=1in,right=1in]{geometry}
\usepackage[sort&compress, numbers, super]{natbib} 
\usepackage{amsfonts, amsmath, amssymb, amsthm, constants, bbm}
\usepackage{MnSymbol}
\usepackage{multirow,lscape}
\usepackage{authblk}

\usepackage{color}
\definecolor{darkred}{RGB}{100,0,0}
\definecolor{darkgreen}{RGB}{0,100,0}
\definecolor{darkblue}{RGB}{0,0,150}

\usepackage{hyperref}
\hypersetup{colorlinks=true, linkcolor=darkred, citecolor=darkgreen, urlcolor=darkblue}
\usepackage{url}

\newcommand{\secref}[1]{Section~\ref{sec:#1}}
\newcommand{\figref}[1]{Figure~\ref{fig:#1}}
\newcommand{\tabref}[1]{Table~\ref{tab:#1}}

\newcommand{\ps}{\mbox{PS}}

\usepackage{setspace}
\begin{document}
\thispagestyle{empty}

\providecommand{\keywords}[1]
{
  \textbf{\textit{Keywords}} #1
}

\title{Sensitivity Analysis of Treatment Effect to Unmeasured Confounding in Observational Studies with Survival and Competing Risks Outcomes}
\author[1]{Rong Huang}
\author[1,2]{Ronghui Xu}
\author[3]{Parambir S. Dulai}
\affil[1]{Department of Mathematics, University of California, San Diego}
\affil[2]{Department of Family Medicine and Public Health, University of California, San Diego}
\affil[3]{Department of Medicine, University of California, San Diego}

\renewcommand\Authands{ and }
\date{}

\maketitle

\begin{abstract}
    No unmeasured confounding is often assumed in estimating  treatment effects in observational data when using approaches such as propensity scores and inverse probability weighting. However, in many such studies due to the limitation of the databases, collected confounders are not exhaustive, and it is crucial to examine the extent to which the resulting estimate is sensitive to the  unmeasured confounders. We consider this problem for survival and competing risks data. Due to the complexity of models for such data, we adapt the simulated potential confounders approach of Carnegie {\it et al.}~(2016), which provides a general tool for sensitivity analysis due to unmeasured confounding. More specifically, we specify one sensitivity parameter to quantify the association between an unmeasured confounder and the treatment assignment, and another set of parameters to quantify the association between the confounder and the time-to-event outcomes. By varying the magnitudes of the sensitivity parameters, we estimate the treatment effect of interest using the stochastic EM and the EM algorithms. We demonstrate the performance of our methods on  simulated data, and apply them to a comparative effectiveness study in  inflammatory bowel disease (IBD). 
\end{abstract}


\section{Introduction}

When estimating the effect of a treatment or exposure on an outcome of interest, the gold-standard approach is to conduct randomized control trials (RCT). In this setting, the potential outcomes are independent of treatment assignment, then the inference is straightforward. However, in many cases, RCT is not feasible and the inference has to be drawn from observational studies. One of the major challenges in inferring causality from observational studies is that the treatment selection mechanism is unknown and we have to rely on some untestable assumptions. One widely used assumption is that there is no unobserved confounding, which means that the treatment assignment and the outcome are independent conditional on the observed pre-treatment covariates. In the potential outcome framework and under this assumption, there are several methods to adjust for the pre-treatment covariates via the propensity, including matching, stratification, outcomes regression and inverse probability weighting (IPW).\cite{RosenbaumRubin83,RosenbaumOS,Austin11, austin2014use, 
d1998propensity}

Instead of believing that the no-unobserved-confounding assumption is satisfied, sensitivity analysis offers an 
approach to assess the extent to which the inference is robust to potential unmeasured confounders. \cite{RosenbaumOS}
Robins {\it et al.}\cite{robins2000sensitivity} argued that sensitivity analysis should be conducted to examine how the inference varies if any untestable assumption is violated. 
 For example, an early sensitivity analysis was conducted in Cornfield {\it et al.}, \cite{cornfield1959smoking} who concluded that the association between cigarette smoking and lung cancer could be explained away only if there existed a hidden bias associated with cigarette smoking that was at least as strong as the association between cigarette use and lung cancer.
Rosenbaum\cite{RosenbaumOS} contains a nice introduction  describing the idea based on association between the unobserved confounder and the treatment and between the unobserved confounder and the outcome.  
Analytical approaches have been developed for simpler outcomes such as binary.\cite{liu2013introduction} 
Li {\it et al.}\cite{li2011propensity} and Shen {\it et al.}\cite{shen2011sensitivity} considered
sensitivity analysis methods for inverse probability weighted (IPW) estimators using propensity scores that were gaining popularity in practice.\cite{AustinStuart15}

Our motivation came from studies in inflammatory bowel disease (IBD). IBD is an umbrella term for two conditions, ulcerative colitis (UC) and Crohn’s disease (CD), that are characterized by chronic inflammation of the gastrointestinal tract. \cite{katz2011facts}  While randomized clinical trials exist, these RCTs only represent 30\% of patients seen and cared for in routine practice. Furthermore, there are a growing number of treatment options, and head-to-head comparisons are entirely lacking due to difficulty in performing these RCTs and the rapid growth in treatment options. 
In order to compare the effectiveness between Vedolizumab and tumor necrosis factor (TNF)-antagonist therapies for UC and CD patients, data were collected between May 2014 and December 2017 from a North American based consortium registry,\cite{narula2018vedolizumab,dulai2016real} which is a multi-center collaborative research group where outcomes are pooled for consecutive UC and CD patients treated with biologics. Our primary endpoint is time to clinical remission. 
Although data collection was rather exhaustive and accounted for most known measurable confounders, treatment selection for IBD is known to be preference sensitive and influenced by patient and provider perceptions, experiences, and understandings of potential benefit and risk based on the data available to them, all of which are unmeasurable. 
We aim to assess to what extent our inference from the data is affected by potentially unmeasured confounding. 

Time to clinical remission is a survival endpoint;
however, patients need time to achieve this endpoint. Wide variability exists across centers, patients, and providers, for their preference to proceed with surgery while awaiting response to therapy.  Therefore, surgery presents a competing risk to clinical remission, in that surgery prevents the event of achieving clinical remission.
In Lukin {\it et al.}\cite{dulai_UC} and Bohm {\it et al.}\cite{dulai_CD} the authors considered propensity score methods with IPW as the primary approach to account for the observed covariates. However, it is possible that there might be confounders not captured by the observed covariates. To carry out sensitivity analysis for this type of complex outcomes, we found the simulated unobserved confounder approach\cite{carnegie2016assessing} to be useful, since the analytical approaches seem difficult to derive. 


Under the simulated unobserved confounder setting we may consider  two types of sensitivity parameters, one describing the association between the unmeasured confounder and the treatment assignment and the other  describing the association between the unmeasured confounder and the outcome. The interpretation of these parameters is relatively straightforward.
To determine a plausible range of these sensitivity parameters, we take into consideration the maximum observed association between a measured confounder and the treatment assignment or the outcome, as well as the typical consideration of the strength of association such as the odds ratio or hazard ratio for (say) a binary predictor, as we will see later in the paper.


We organize our paper as follows. We describe our models in \secref{models}, including both the survival models and the competing risks models. We consider estimation in \secref{algo},  using both the Expectation-Maximization (EM) algorithms and a stochastic EM algorithm.
 In \secref{simulation}, we demonstrate the performance of our algorithms via simulations. We  apply our methods to the IBD data in \secref{IBD}. Finally, we conclude  with  discussion in \secref{discussion}.

\section{Models} \label{sec:models}


\subsection{Survival outcome}

Denote $T^0$  a time-to-event outcome,  $Z$ a binary treatment assignment, and  $X$ a vector of covariates. 
Due to possible right censoring, we observe $T=\min(T^0, C)$ and $\delta=I(T^0 \leq C)$, where $C$ is the censoring time random variable, and $I(\cdot)$ the indicator function.
We consider $U$  which represents the portion of  unmeasured confounder(s) that is independent of $X$, and will simply refer to $U$ as the unmeasured confounder for the rest of the paper. 
We assume $U$ to be binary for ease of implementation, although other distributions are possible and will be discussed later.  
Given $Z$, $X$ and $U$, the hazard rate of $T^0$ 
    is modeled using the 
 Cox proportional hazards (PH) regression:\cite{cox1972regression}
\begin{equation}\label{model:surv}
    \lambda (t| Z, X, U) = \lambda_{0} (t) \exp(\tau Z + \beta' X+ \zeta U).  
\end{equation} 
 In addition, we assume that given $X$ and $U$, $Z$ follows a generalized linear model; for illustration purposes we assume a probit link below, although logistic would be an obvious alternative:
\begin{equation}\label{model:ps}
     P( Z=1 |X,U ) = \Phi(X' \beta^z + \zeta^z U),
\end{equation} 
where $\Phi$ is the standard normal cumulative distribution function (CDF).
In the above $(\zeta^z, \zeta)$ are  sensitivity parameters, which quantify the relationships between the unobserved confounder and the treatment assignment and the outcome, respectively.
Finally, we assume that $U \sim$ Bernoulli$(\pi)$, and we set $\pi=0.5$. 

Our goal is to simulate $U$ given the observed $T$, $\delta$, $Z$ and $X$. We note that  if the parameters in the above are known, then 
\begin{equation} \label{conU}
    U|T,\delta, Z,X \sim \mbox{Bernoulli} \left(\frac{\pi^{T, \delta, Z, X, U=1}}{\pi^{T, \delta, Z, X}} \right),
\end{equation} 
where $\pi^{T, \delta, Z, X, U=u}$ is the joint likelihood of $(T, \delta, Z, U=u)$ given $X$ for $u=0, 1$, and $\pi^{T, \delta, Z, X} = \pi^{T, \delta, Z, X, U=1} + \pi^{T, \delta, Z, X, U=0}$. 
In particular, 
\begin{eqnarray} \label{jointlik}
    \pi^{T,\delta, Z,X,U} &=& \pi^U (1-\pi)^{1-U} \left\{\Phi(X' \beta^z + \zeta^z U) \right\}^Z \left\{1-\Phi(X' \beta^z + \zeta^z U) \right\}^{1-Z} \nonumber \\
    &&\cdot \left\{ \lambda_{0} (T) e^{ \tau Z + \beta' X + \zeta U} \right\}^\delta  \exp \left\{ - \Lambda_{0} (T) \cdot e^{  \tau Z +  \beta' X + \zeta U} \right\}.
\end{eqnarray}

\subsection{Competing risks}

In the presence of competing risks, when an event occurs it may be one of $m$ distinct types of failures indexed by $j = 1, 2, \cdots, m$. 
Denote $T^1, \cdots, T^m$ the potential time-to-event outcomes for the $m$ types, and as before $Z$ a binary treatment assignment, and $X$ a vector of covariates. We observe $T=\min(T^1, \cdots, T^m, C)$, where $C$ is the censoring time random variable, $\delta=I(\min(T^1, \cdots, T^m) \leq C)$ and $J$ the type of failure if not censored. Again, we consider an unmeasured binary confounder $U$ that is independent of $X$. 
The cause-specific hazard function\cite{kalbfleisch2011statistical} for the $j$-th failure type is 
  $  \lambda_j (t| Z,X,U) = \lim_{\Delta t \to 0}  \mathbb{P}(t \le T < t+\Delta t, J = j|T \ge t, Z,X,U) / {\Delta t}$.
We consider the proportional cause-specific hazards model (PCSH)
\begin{equation}\label{model:comprisk}
    \lambda_j (t| Z,X,U) = \lambda_{j0} (t) \exp( \tau_j Z + \beta'_j X+ \zeta_j U), \quad j = 1,2, \cdots, m.
\end{equation} 
As before we also assume that given $X$ and $U$, $Z$ follows a generalized linear model \eqref{model:ps} with a probit link. Then parallel to \eqref{jointlik} we have
\begin{align}
    \pi^{T,\delta, J,Z,X,U} &= \pi^U (1-\pi)^{1-U} \left\{\Phi(X' \beta^z + \zeta^zU) \right\}^Z \left\{1-\Phi(X' \beta^z + \zeta^zU) \right\}^{1-Z}  \nonumber \\
    &\cdot \prod_{j=1}^m  \left\{ \lambda_{j0} (T) e^{ \tau_j Z +  \beta_j' X+ \zeta_j U} \right\}^{I(\delta=1, J=j)}  \exp \left\{- \Lambda_{j0} (T) \cdot e^{\tau_j Z +  \beta_j' X+ \zeta_j U} \right\},
\end{align}
where {$I(\delta=1, J=j)$} indicates whether subject had the event $j$. The posterior probability of $U$ is then obtained similar to \eqref{conU}. 
In general, if there are $m$ distinct types of failures, then there would be $m+1$ sensitivity parameters, $\zeta^z, \zeta_1, \cdots, \zeta_m$.

\section{Estimation} \label{sec:algo}

In order to simulation $U$ given the observed data, we first need to estimate the unknown parameters. 
Conditional on the unobserved $U$ as well as $Z$ and $X$, the likelihood function of the survival outcome without competing risks is 
\begin{align}
	L_1 &= \prod_{i=1}^n \lambda (t_i| z_i, x_i, u_i)^{\delta_i} \exp \{-\Lambda (t_i| z_i, x_i, u_i) \} \nonumber\\
	&= \prod_{i=1}^n \left\{ \lambda_{0} (t_i) e^{\tau z_i + \beta' x_i+ \zeta u_i} \right\}^{\delta_i}  \exp \left\{-\Lambda_{0} (t_i) e^{\tau z_i + \beta' x_i+ \zeta u_i} \right\}, 
\end{align}
Similarly, the likelihood function of the competing risks outcome given $(Z, X, U)$   is 
\begin{eqnarray}\label{comp_lik}
L_1 &=& \prod_{i=1}^n \prod_{j=1}^m \lambda_j (t_i| z_i, x_i, u_i)^{\delta_{ij}} \exp \{-\Lambda_j (t_i| z_i, x_i, u_i)\} \nonumber\\
&=& \prod_{i=1}^n \prod_{j=1}^m \left\{ \lambda_{j0} (t_i) e^{ \tau_j z_i + \beta'_j x_i+ \zeta_j u_i} \right\}^{\delta_{ij}}  \exp \left\{-\Lambda_{j0} (t_i) e^{ \tau_j z_i + \beta'_j x_i+ \zeta_j u_i} \right\}, 
\end{eqnarray}
{where $\delta_{ij} := I(\delta_i=1, J_i=j)$ indicates whether subject $i$ had event $j$. }

\subsection{The EM algorithm}

The EM algorithm\cite{dempster1977maximum} is a commonly used approach to handle missing data, in this case $U$,  in the likelihood function.  Let $\theta$ 
denote the unknown parameters, 
and $y_i$ 
 the survival outcome for subject $i$. 
The EM algorithm iterates between the E-steps and the M-steps that are described below, where in the notation the covariate $x_i$ is suppressed which is always being conditioned upon. 
The initial values can be set using the parameter estimates from the  regression models ignoring $U$.
We note that the sensitivity parameters, as well as $\pi=0.5$, are known. 

\subsubsection*{E-step}

In the E-step we compute the conditional expectation of the log-likelihood of the complete data $(y_i,z_i, u_i)$ given the observed data and the current  parameter value $\tilde \theta$. For the survival outcome without competing risks, let
\begin{align}\label{Q}
    \mathcal{Q}(\theta) &= \mathbb{E}[l(\theta;\mathbf{y},\mathbf{z},\mathbf{u})|\mathbf{y},\mathbf{z},\tilde \theta] \nonumber\\
    &= \mathbb{E}[l_1(\beta, \tau, \lambda_{0};\mathbf{y}|\mathbf{z},\mathbf{u})|\mathbf{y},\mathbf{z},\tilde \theta] + \mathbb{E}[l_2(\beta^z;\mathbf{z}|\mathbf{u})|\mathbf{y},\mathbf{z},\tilde \theta] + \mathbb{E}[l_3(\mathbf{u})|\mathbf{y},\mathbf{z},\tilde \theta]\\
    &:= \mathcal{Q}_1(\beta, \tau, \lambda_{0}) + \mathcal{Q}_2 (\beta^z) + \mathcal{Q}_3, \nonumber
\end{align} 
where
\begin{align} \label{Q1}
    \mathcal{Q}_1(\beta, \tau, \lambda_{0}) &= \sum_{i=1}^n 
    \left[ \delta_{i} \{ \log \lambda_{0} (t_i) + \beta' x_i+ \zeta \mathbb{E}[u_i|y_i, z_i, \tilde \theta] + \tau z_i \} \right. \nonumber\\
    & \left. \quad -\Lambda_{0} (t_i) \exp \left\{ \beta' x_i+ \log \mathbb{E}[e^{\zeta u_i}|y_i, z_i, \tilde \theta] + \tau z_i \right\} \right],
\end{align}
\begin{equation} \label{Q2}
\mathcal{Q}_2 (\beta^z) = \sum_{i=1}^n \{z_i\mathbb{E}[\log (\Phi(x_i' \beta^z + \zeta^z u_i))|y_i, z_i, \tilde \theta] + (1-z_i) \mathbb{E}[\log  (1-\Phi(x_i' \beta^z + \zeta^z u_i))|y_i, z_i, \tilde \theta]\},
\end{equation}
\begin{equation} \label{Q3}
\mathcal{Q}_3 = \sum_{i=1}^n \{\log \pi \mathbb{E}[u_i|y_i, z_i, \tilde \theta] + \log (1-\pi) \mathbb{E}[1-u_i|y_i, z_i, \tilde \theta] \}.
\end{equation} 
We note that $\mathcal{Q}_3$ is in fact not used in the M-step since it does not involve unknown parameters. 
As described earlier, given the observed data, $U$ follows Bernoulli $(\tilde \pi_i)$ as in \eqref{conU} where 
$ \tilde \pi_i $ is calculated based on the current parameter value $\tilde \theta$. 
So for any function $h(u_i)$ in \eqref{Q1} and \eqref{Q2}, we have
$\mathbb{E}[h(u_i)|y_i, z_i, \tilde \theta] 
= h(1) \tilde \pi_i + h(0) (1-\tilde \pi_i)$.

For competing risks outcome, from \eqref{comp_lik} we see that 
 the likelihood function  is a product of $m$ likelihoods, one for each type of event with its own type specific parameters. The corresponding $\mathcal{Q}_1$ function is then a sum of $\mathcal{Q}_{1j}(\beta_j, \tau_j, \lambda_{j0})$'s, each having the same form as $\mathcal{Q}_1(\beta, \tau, \lambda_{0})$ above but with parameters $ \beta_j, \tau_j, \lambda_{j0} $ and data for the event type $j$ instead.

\subsubsection*{M-step}

From (\ref{Q}) it is clear that in 
the M-step we can update $(\beta, \tau, \lambda_{0})$ and $\beta^z$ separately. In order to maximize $\mathcal{Q}_1$, we note that it has the same form as the log-likelihood in a Cox regression model with known offset $\log \mathbb{E}[e^{\zeta u_i}|y_i, z_i, \tilde \theta]$, just like the Cox model with random effects.\cite{vaida2000proportional} 
For competing risks again because $\mathcal{Q}_1$  is  a sum of $\mathcal{Q}_{1j}(\beta_j, \tau_j, \lambda_{j0})$'s for $j=1, ..., m$, each set of parameters $ \beta_j, \tau_j, \lambda_{j0} $ is updated separately using the Cox model software with offsets, the same way as a single survival outcome.

To maximize $\mathcal{Q}_2$, we have
\begin{align}
    \mathcal{Q}_2 (\beta^z) &= \sum_{i=1}^n \left(z_i \left[\log \{\Phi(x_i' \beta^z + \zeta^z)\}\tilde \pi_i + \log \{\Phi(x_i' \beta^z)\} (1-\tilde \pi_i) \right] \right. \nonumber \\
    & \left. \quad + (1-z_i) \left[\log  \{1-\Phi(x_i' \beta^z + \zeta^z)\}\tilde \pi_i + \log  \{1-\Phi(x_i' \beta^z)\}(1-\tilde \pi_i) \right]\right). 
\end{align}
This function can be maximized using the R function `{optim}'.

\subsubsection*{Variance estimation}

As in typical nonparametric maximum likelihood inference under semiparametric models, 
the variance-covariance matrix of $\hat \theta$ is estimated by the inverse of a discrete observed information matrix $I(\hat \theta)$ following the EM algorithm, which is given by Louis' formula \cite{louis1982finding} based on missing information principle: 
\begin{equation} \label{info}
I(\theta) = \mathbb{E}[-\ddot{l}(\theta;\mathbf{y},\mathbf{z},\mathbf{u})|\mathbf{y},\mathbf{z},\theta] - \mathbb{E}[s(\theta;\mathbf{y},\mathbf{z},\mathbf{u})s(\theta;\mathbf{y},\mathbf{z},\mathbf{u})'|\mathbf{y},\mathbf{z},\theta],
\end{equation}
where $\ddot{l}$ and $s$ denote the second and first derivatives of $l$ with respect to $\theta$. 
The components of $\ddot{l}$ and $s$ are given in the Appendix.

\subsection{The Stochastic EM algorithm}

Instead of the EM algorithm described above, the stochastic EM algorithm was used in Carnegie {\it et al.} \cite{carnegie2016assessing}, we think primarily due to its ease of implementation for practitioners as well as intuitive appeal. It is similar to a Monte Carlo EM (MCEM) but in the E-steps only a single $U$ is drawn from the conditional distribution of $U$ given the observed data, so that in the M-steps the parameters are updated using that single sample of $U$ as if it were observed. A typical MCEM would otherwise draw many samples of $U$ in order to approximate the conditional expectations in the E-steps. The E- and M-steps are as described above for the models that we consider in this paper, for both survival and competing risks outcomes.
 
At the convergence of the stochastic EM algorithm, in order to obtain a more accurate estimate the E- and M-steps are  repeated  $K$ times, 
 and the final estimate of the treatment effect on the survival outcome is 
  $\hat \tau =  \sum_{k=1}^K \hat \tau_{k} /K$, with the corresponding standard error 
\begin{equation}\label{mi_var}
    \hat \sigma_{\hat \tau} = \sqrt{\frac{1}{K} \sum_{k=1}^K \hat \sigma^2_{\hat \tau_{k}} + \frac{1}{K-1} \sum_{k=1}^K (\hat \tau_{k} - \hat \tau )^2},
\end{equation} 
where $ \hat \sigma^2_{\hat \tau_{k}} $ is estimated variance of $ \hat \tau_{k} $ pretending that the singly sampled $U_k$ is observed. 
For competing risks we have similarly
for  type $j$ event  
 $\hat \tau_j =  \sum_{k=1}^K \hat \tau_{jk} /K$, and the corresponding standard error is obtained using \eqref{mi_var} with $ \hat \tau_{k} $ replaced by $ \hat \tau_{jk} $ and $\hat \tau$ replaced by $\hat \tau_j$. 

Nielsen {\it et al.}\cite{nielsen2000stochastic} studied the asymptotic behavior of the stochastic EM algorithm, and showed that under certain assumptions it is root-$n$ consistent but not fully efficient. 
We show in our data analysis that it can be naturally adapted to the IPW approach and obtain inferential results in sensitivity analysis. 

\section{Simulations} \label{sec:simulation}

\begin{figure}[!thpb]
	\centering	
	\subfigure[$\hat \tau_1$]{
		\includegraphics[width=0.45\textwidth]{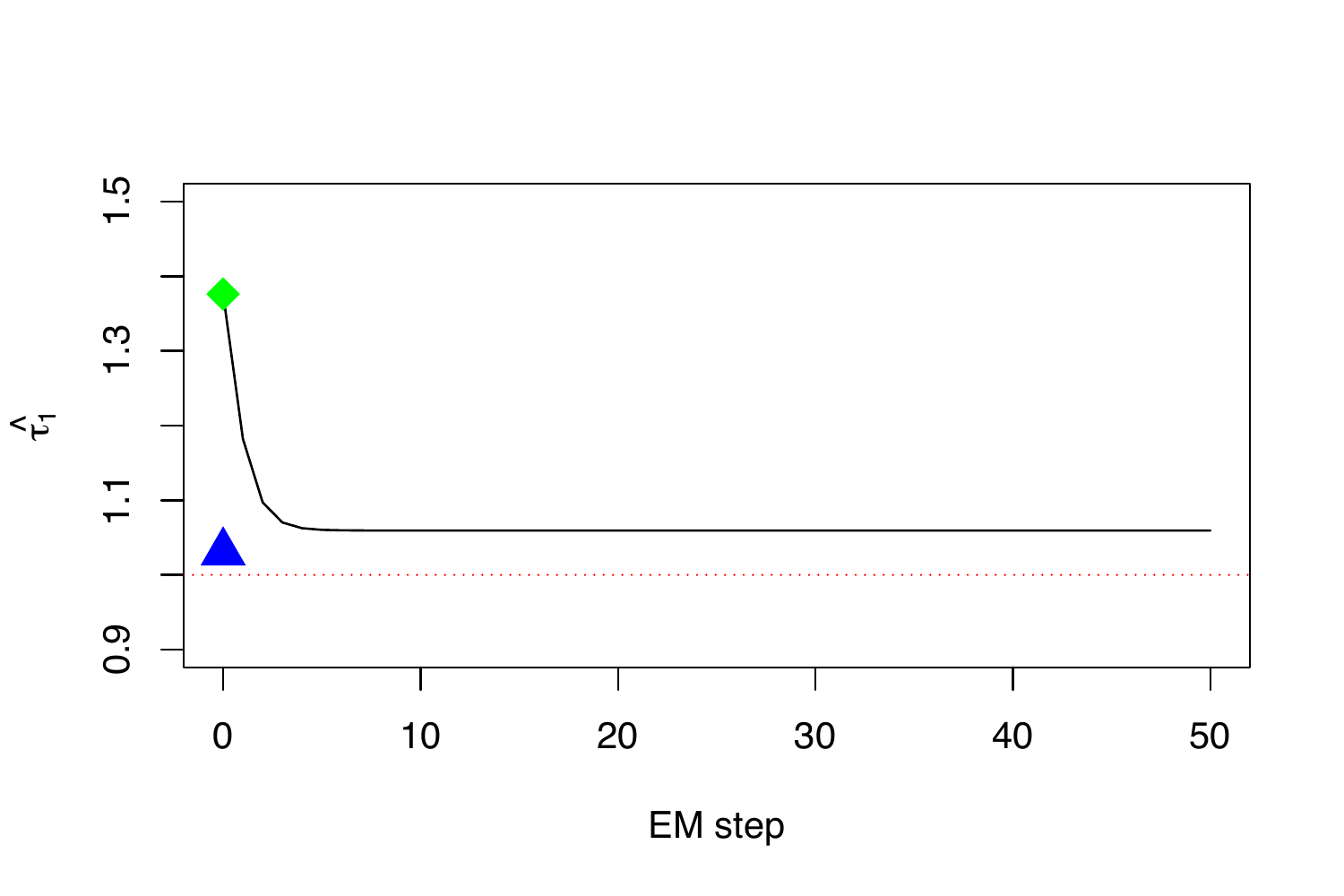}
		\label{fig:tau1EM}
	}	
	\subfigure[$\hat \tau_2$]{
		\includegraphics[width=0.45\textwidth]{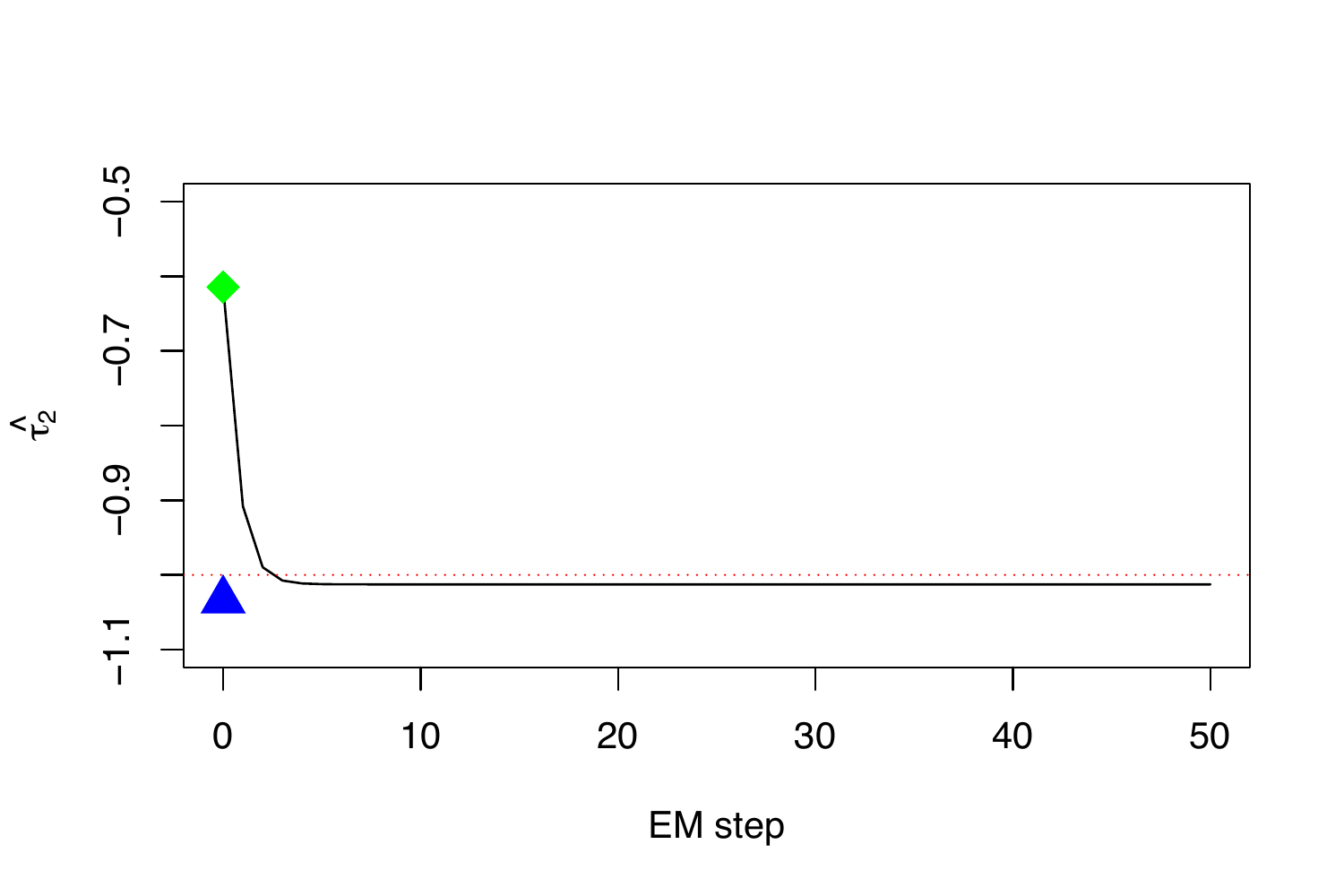}
		\label{fig:tau2EM}
	} \\
	\subfigure[$\hat \tau_1$]{
		\includegraphics[width=0.45\textwidth]{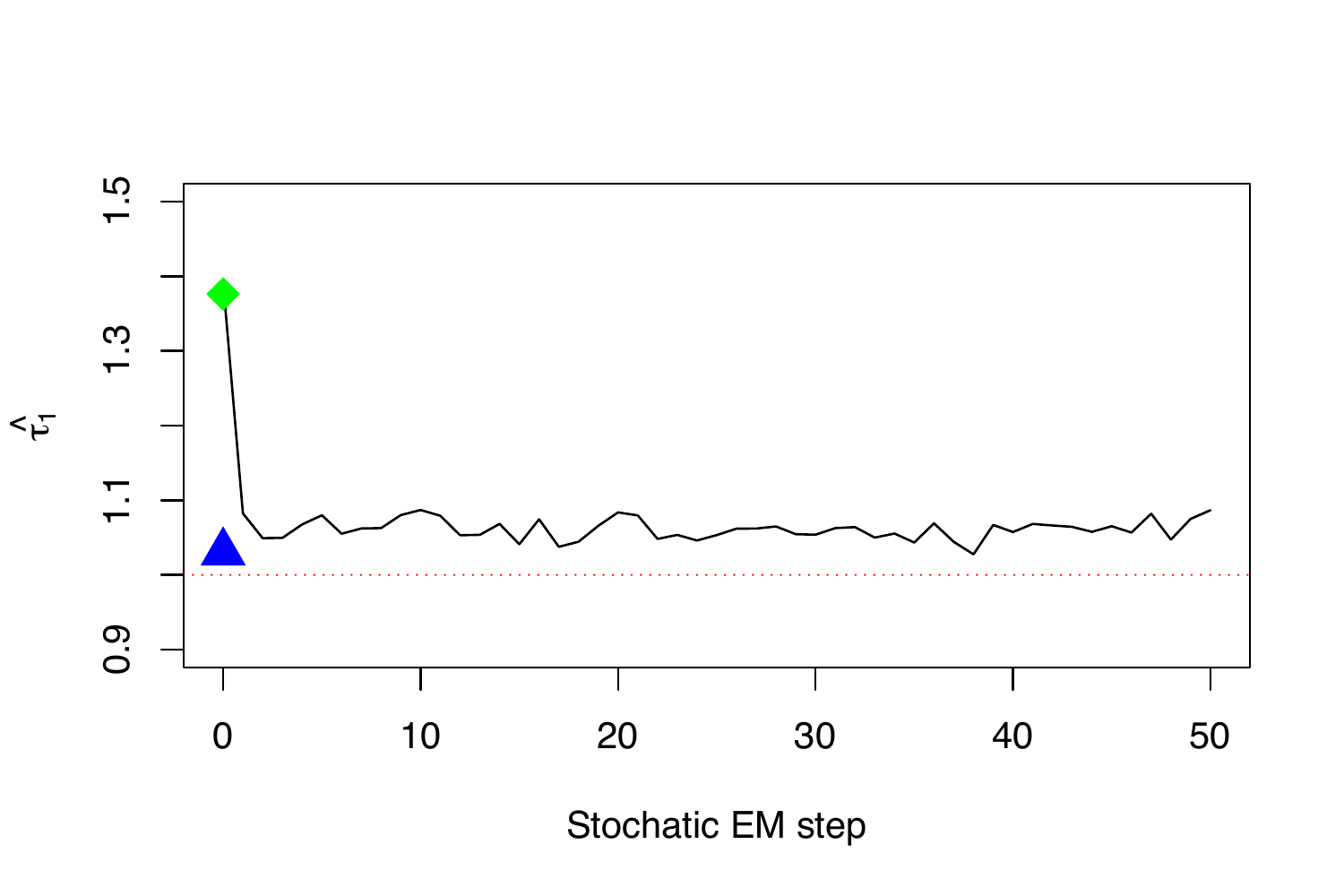}
		\label{fig:tau1offset}
	}	
	\subfigure[$\hat \tau_2$]{
		\includegraphics[width=0.45\textwidth]{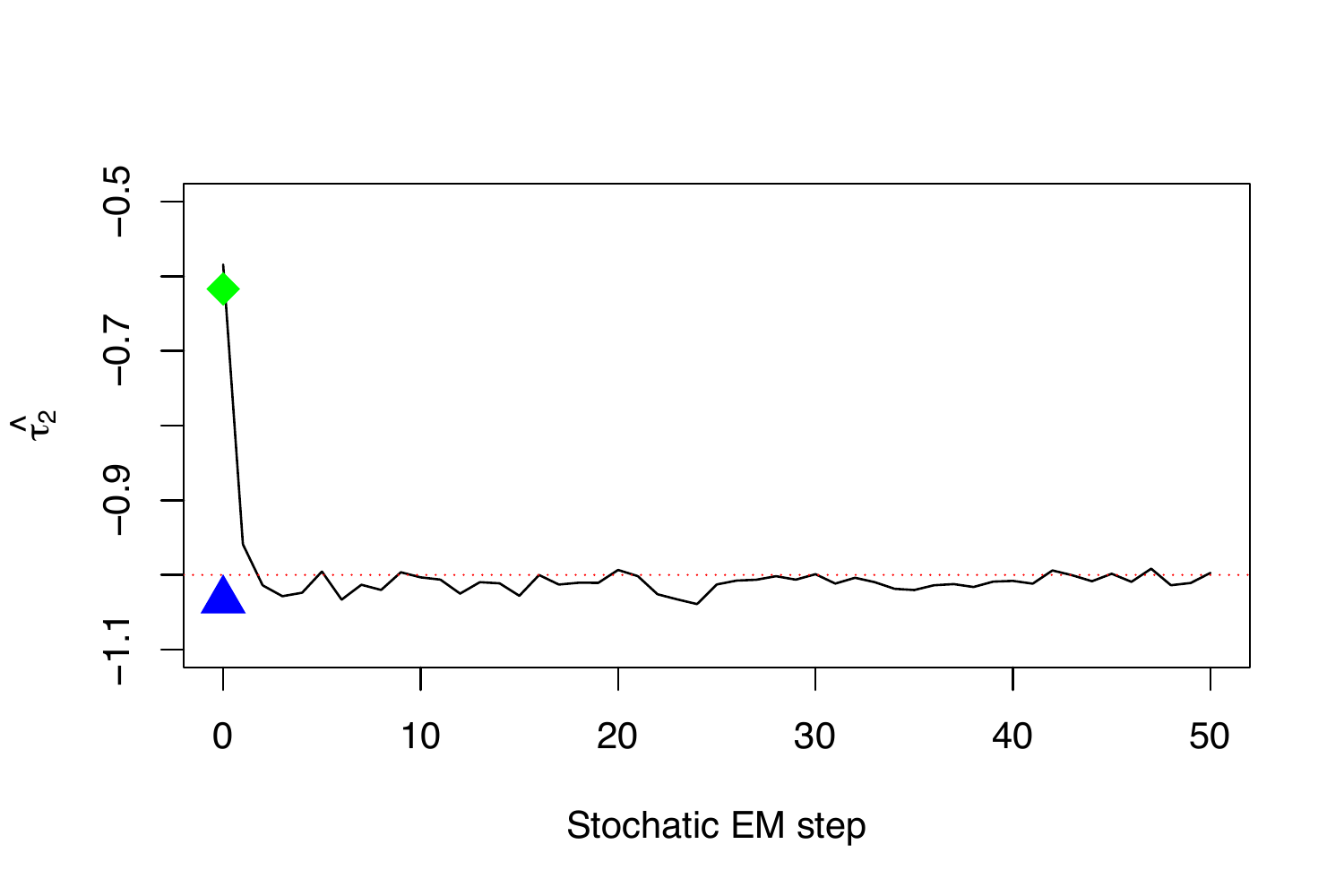}
		\label{fig:tau2offset} 
	}
	\caption{Convergence of the EM (top) and stochastic EM (bottom) algorithms in a single simulation run. 
	The blue triangles correspond to the estimated treatment effect with true $U$ and the green diamonds correspond to the estimates without $U$. The red horizontal lines indicate the true values of $\tau_j$'s, and $\zeta_1 = \zeta_2 = \zeta^z = 1$. }
	\label{fig:EM}
\end{figure}

We conducted  simulation studies to investigate the performance of the EM as well as the stochastic EM algorithms, as compared to the estimation of the treatment effect using the true confounder $U$ with the given 
 sensitivity parameters. For both survival and competing risks outcomes, we set 
 sample size $n=1,000$, $U \sim$ Bernoulli(0.5), and two independent covariates $X_1 \sim N(0,1)$, $X_2 \sim N(1,1)$ with 
$ \beta^z = (0.25, -0.25)' $ in \eqref{model:ps}. 
The number of EM or stochastic EM steps was set to 20 (see \figref{EM} and related discussion below), and true sensitivity parameter values were used in fitting the models. 
The final estimates from the stochastic EM were obtained by averaging over $K=40$ estimates to reduce the variability. 
For each case we show the results of 200 simulation runs.

\subsection{Survival outcome}

To simulate survival outcomes under  model \eqref{model:surv}, we set $\lambda_{0}(t) =1$, $\beta = (0.5, -1)'$ and $\tau = 1$. 
In addition, we set censoring times $C\sim$ Uniform(1, 2) which led to between 25$\sim$60\% censoring, depending on the combinations of the parameter values.  

\begin{figure}[!thpb]
\centering
\includegraphics[width=1.0\textwidth]{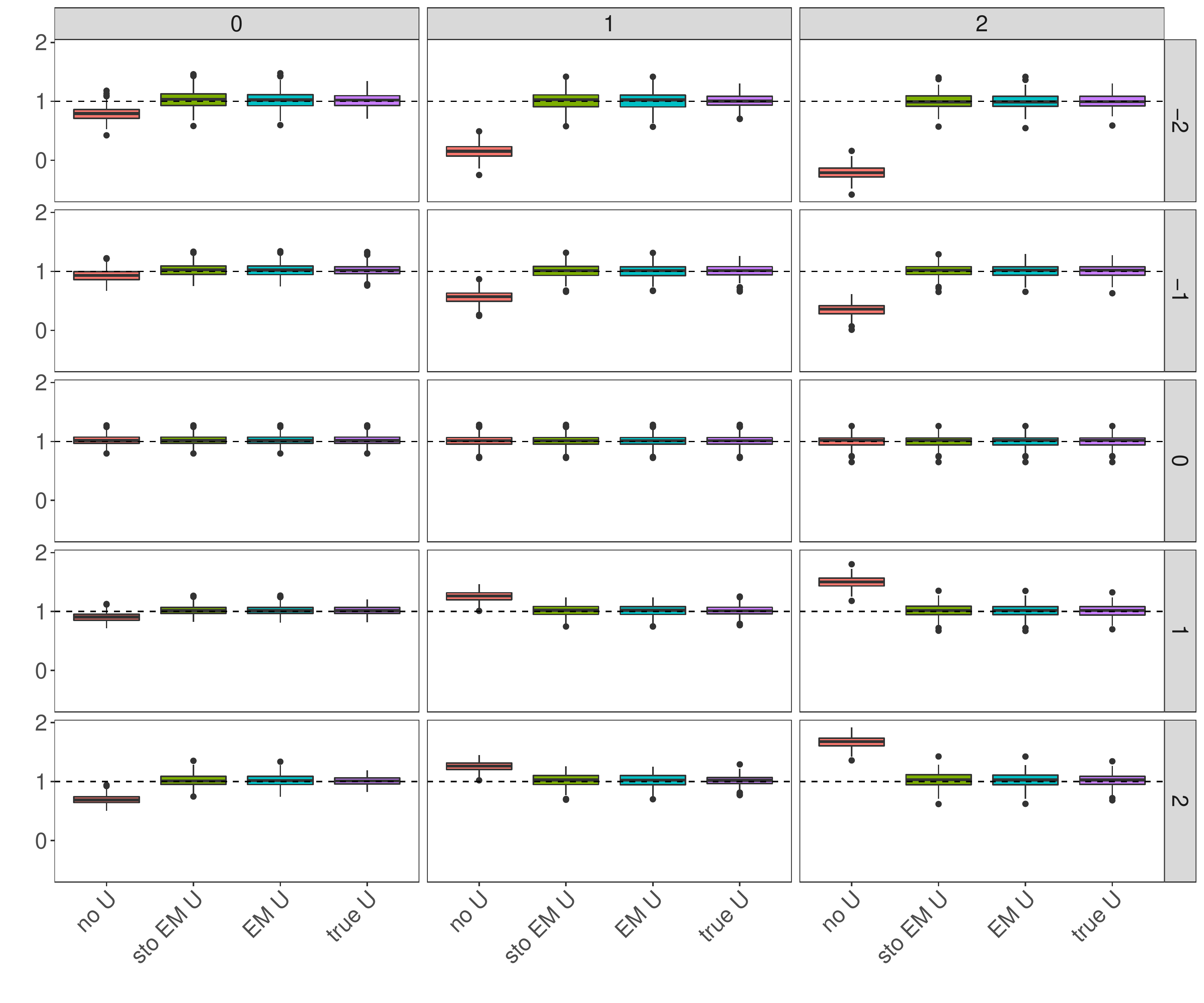}
\caption{Distributions of the estimated treatment effect ($\hat \tau$) for the simulated survival data. $\zeta^z \in \{0,1,2\}$ on the horizontal label and $\zeta \in \{-2,-1, 0,1,2\}$ on the vertical label. Each boxplot displays $\hat \tau$ from 200 simulations.
 }
\label{fig:tau}
\end{figure}

We run simulations over each combination of $\zeta^z \in \{0,1,2\}$ and $\zeta \in \{-2, -1, 0, 1, 2\}$. 
The results of the simulation are reported in \tabref{tau} and \figref{tau}.
From \figref{tau} it is clear that  ignoring $U$ led to bias in the estimated treatment effect as long as $\zeta\neq 0$; this bias also increases with the magnitude of $\zeta$ as well as the magnitude of $\zeta^z$.  
On the other hand, both the stochastic EM and the EM algorithm gave good estimates of the treatment effect compared with the estimates using the true $U$'s. Closer comparison of the results in  \tabref{tau} shows that the EM algorithm gave more accurate estimates than the stochastic EM algorithm, both in terms of generally less bias and smaller variances.

\subsection{Competing risks outcomes}

To simulate competing risks outcomes, we followed the approach designed in Beyersmann {\it et al.} \cite{beyersmann2009simulating}. We assumed that $m=2$,  the baseline hazard functions for type 1 and type 2 failures to be $\lambda_{10}(t) = \lambda_{20}(t) = 1$, and $\beta_1 = (0.5, -1)'$, $\tau_1 = 1$, $\beta_2 =  (-0.5, 0.2)'$, $\tau_2 = -1$ in model \eqref{model:comprisk}.
 We then simulated the survival times with all-causes hazard $\lambda = \lambda_1 + \lambda_2$, and the cause $J$ was generated from Bernoulli trials with $P(J = 1|Z,X,U) = {\lambda_1}/{(\lambda_1 + \lambda_2)}$. We also set censoring times $C\sim$ Uniform(0.3, 0.7). 

\begin{figure}[!thpb]
\centering
\includegraphics[width=1.0\textwidth]{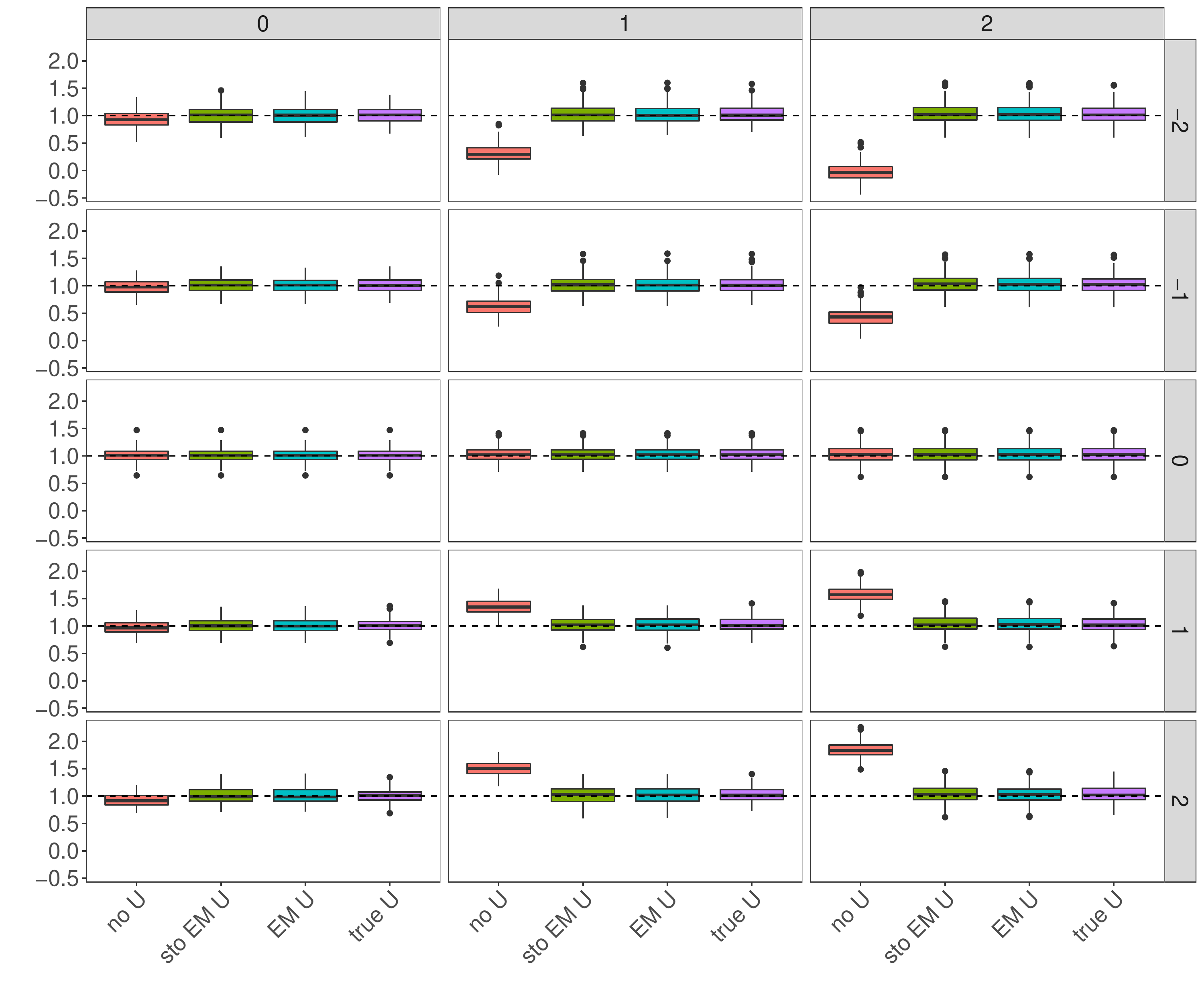}
\caption{Distributions of the estimated treatment effect on type 1 failures for the simulated competing risks data. $\zeta^z \in \{0,1,2\}$ on the horizontal label and $\zeta_1 = \zeta_2 \in \{-2,-1, 0,1,2\}$ on the vertical label. Each boxplot displays $\hat \tau_1$ from 200 simulation runs.}
\label{fig:tau1}
\end{figure}

\begin{figure}[!thpb]
	\centering
	\includegraphics[width=1.0\textwidth]{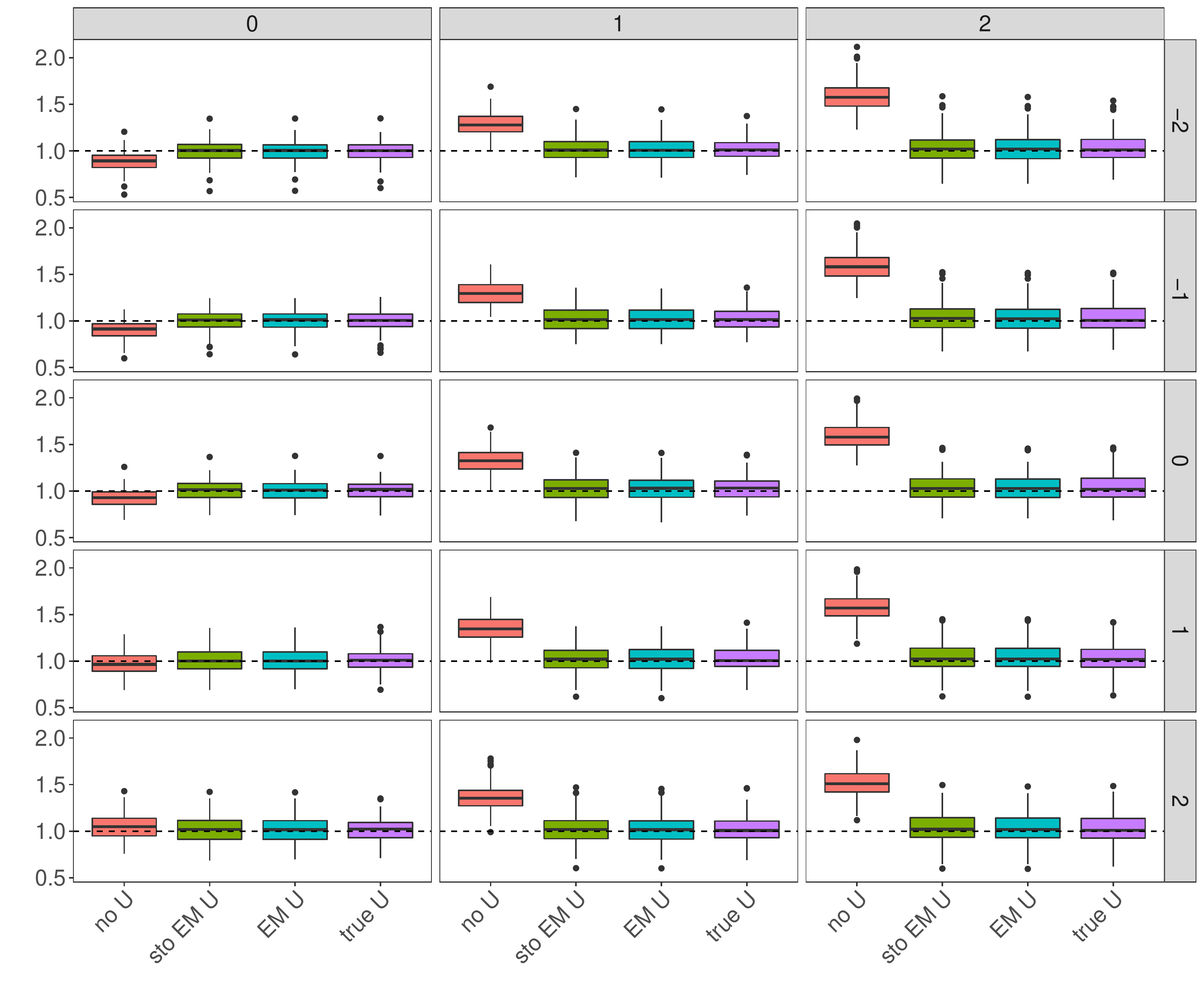}
	\caption{Distributions of the estimated treatment effect on type 1 failures for the simulated competing risks data. $\zeta^z \in \{0,1,2\}$ on the horizontal label, $\zeta_1 =1$ and $\zeta_2 \in \{-2,-1, 0,1,2\}$ on the vertical label. Each boxplot displays $\hat \tau_1$ from 200 simulation runs.}
	\label{fig:tau1_1}
\end{figure}

\begin{figure}[!thpb]
	\centering
	\includegraphics[width=1.0\textwidth]{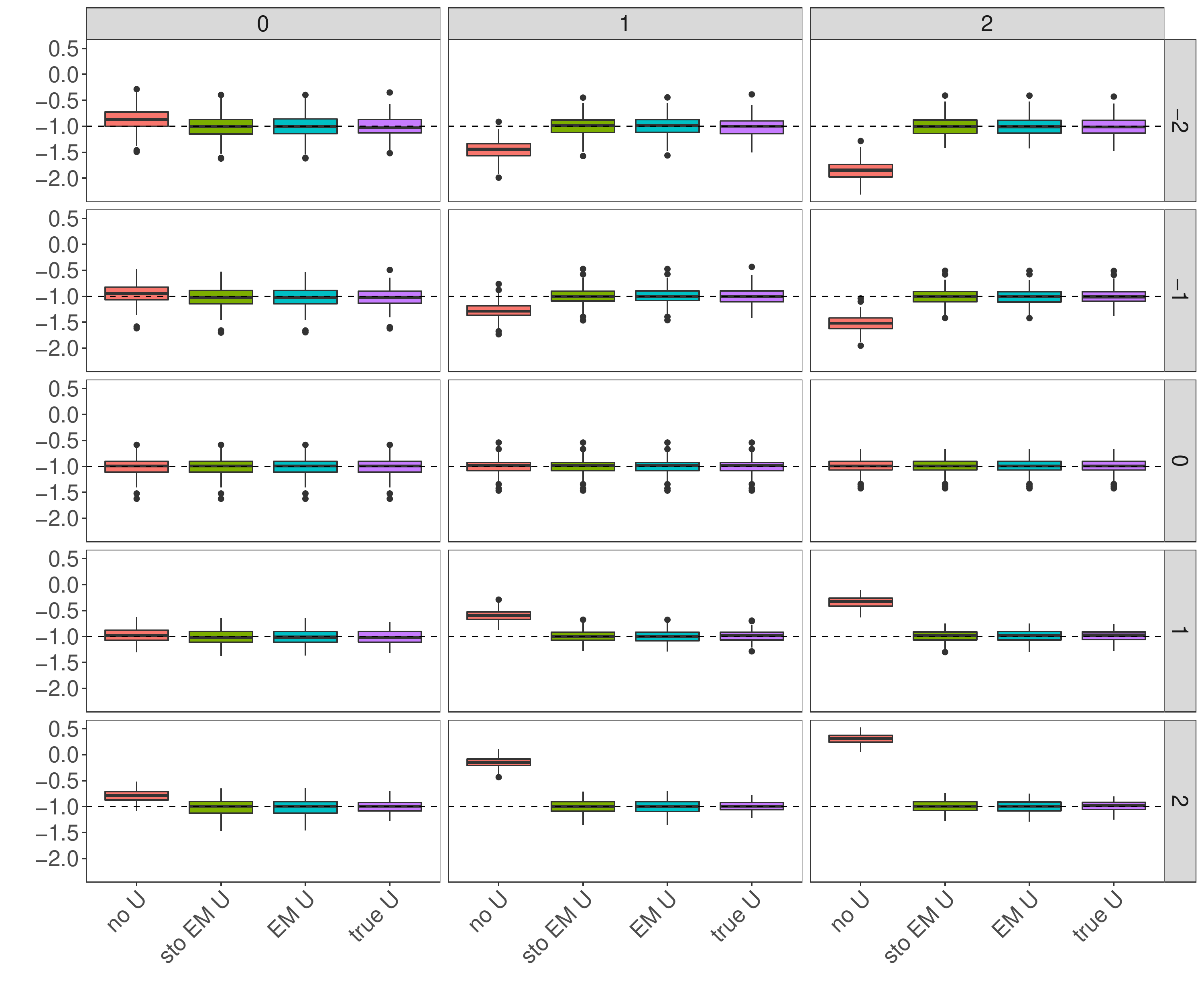}
	\caption{Distributions of the estimated treatment effect on type 2 failures for the simulated competing risks data. $\zeta^z \in \{0,1,2\}$ on the horizontal label, $\zeta_1 =1$ and $\zeta_2 \in \{-2,-1, 0,1,2\}$ on the vertical label. Each boxplot displays $\hat \tau_2$ from 200 simulations. }
	\label{fig:tau2_1}
\end{figure}

Similarly as the survival model, we first ran simulations over each combination of $\zeta^z \in \{0,1,2\}$ and $\zeta_1 =\zeta_2 \in \{-2, -1, 0, 1, 2\}$. This gave about 20$\sim$60\% censoring, depending on the combinations of the parameter values, and about equal numbers of type 1 and type 2 events. In a second scenario, we fixed $\zeta_1 =1$ and $\zeta_2 \in \{-2, -1, 0, 1, 2\}$ as before, which gave about 20$\sim$40\% censoring, and type 1/2 event rates between 40/20\% and 30/50\%, again depending on the combinations of the parameter values. 
The results of experiments are reported in \tabref{tau1} (\figref{tau1}) - \tabref{tau2_1} (\figref{tau2_1}). 
All results show that for each type of failure, the estimated treatment effect by either the stochastic EM or the EM recovered the true treatment effect quite well, while ignoring $U$ induced a substantial bias. In particular, \tabref{tau1_1} and \figref{tau1_1} show that varying $\zeta_2$ had a noticeable impact on the estimation of $\tau_1$, i.e.~unobserved confounding for type 2 failure had a noticeable impact on the estimation of the treatment effect on type 1 failure. 

Finally, we take a closer look at the EM and the stochastic EM algorithm in a single run in \figref{EM}. 
It is seen that the EM sequence  displays a much smoother line than the stochastic EM sequence; and even at convergence, the stochastic EM sequence has quite some fluctuation compared to the EM sequence. 

\section{Sensitivity analysis of the IBD data} \label{sec:IBD}

\subsection{Ulcerative colitis data}

Ulcerative colitis (UC) is one type of IBD that occurs in the large intestine (colon) and the rectum, which is characterized clinically by bloody diarrhea and urgency. We are interested in comparing the effectiveness between Vedolizumab and tumor necrosis factor (TNF)-antagonist therapy for UC patients. The data was collected between May 2014 and December 2017 from the North American based consortium registry.\cite{narula2018vedolizumab} In brief, a total of 719 (453 treated with Vedolizumab, 266 with TNF-antagonist) UC patients with a median follow-up of 12 months were included.
We focus on the treatment effect  of vedolizumab ($Z=1$) versus TNF-antagonist ($Z=0$) on clinical remission, which is defined as resolution of diarrhea, rectal bleeding and urgency. 
 In the Vedolizumab group, 187 patients had clinical remission and no one had surgery, while in the TNF-antagonist group, 100 patients had clinical remission and 3 patients had surgery. 
 Since there were only 3  competing events of surgery, too few to fit any  model, we had to simply treat surgery as independent censoring and applied our approach under the survival models (i.e.~without competing risks) to approximate the treatment effect of Vedolizumab. 
 
In Lukin {\it et al.}\cite{dulai_UC}  the propensity score for each subject $i$, denoted $\ps_i$,
was calculated using the R package `twang'\cite{mccaffrey2004propensity}  based on pre-treatment variables, including age, disease extent, clinical disease severity, UC related hospitalization within the preceding 1-year, prior TNF-antagonist exposure, baseline steroid dependency or refractoriness, concomitant steroid use, and concomitant immunomodulator use. 
Though the above potential confounders were considered, it is unknown if all confounders have been included. Hence, sensitivity analysis is necessary for this data.

\begin{figure}[!thpb]
	\centering	
	\subfigure[stochastic EM]{
		\includegraphics[width=0.45\textwidth]{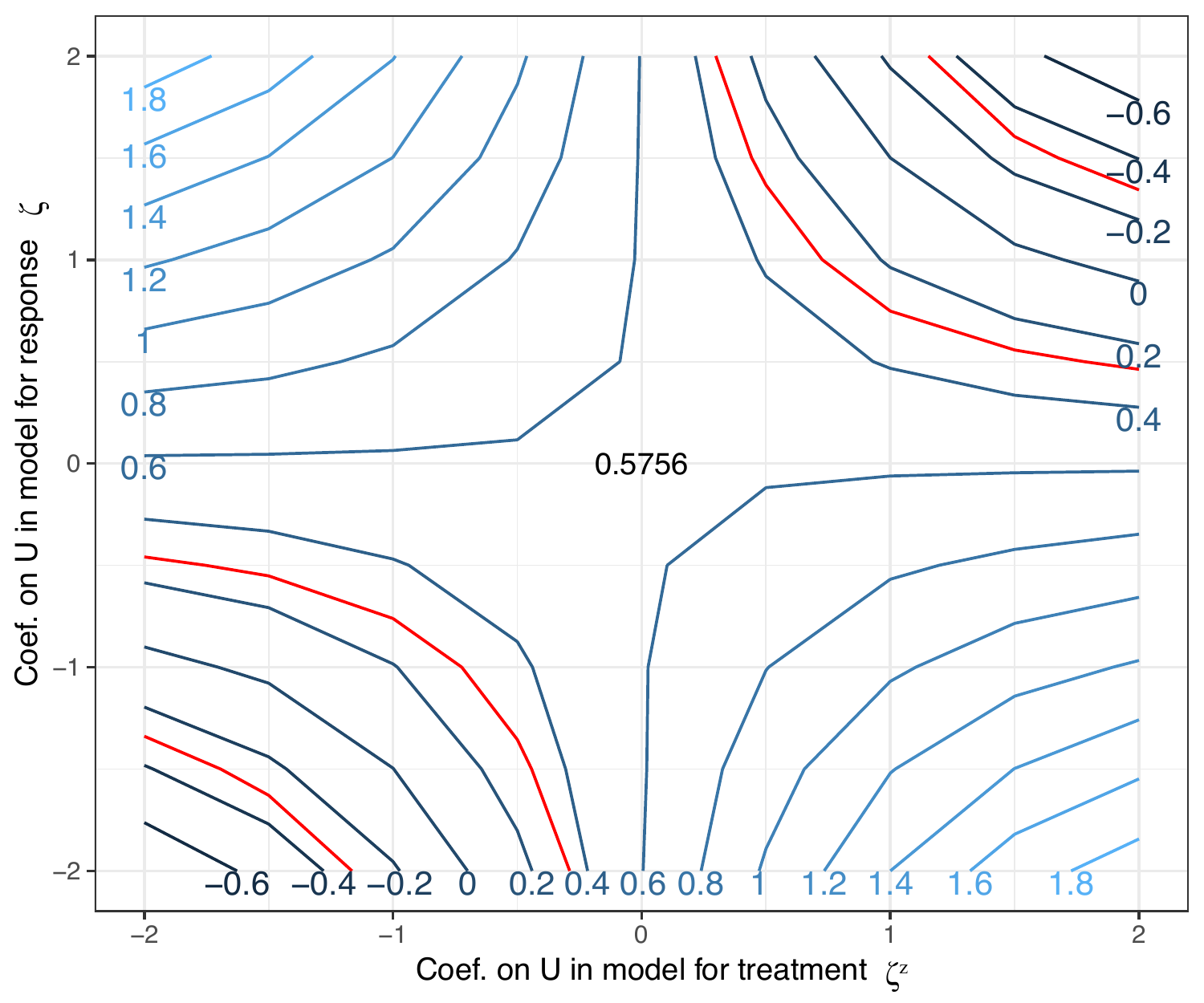}
		\label{fig:UCout1stoEM}
	}	
	\subfigure[EM]{
		\includegraphics[width=0.45\textwidth]{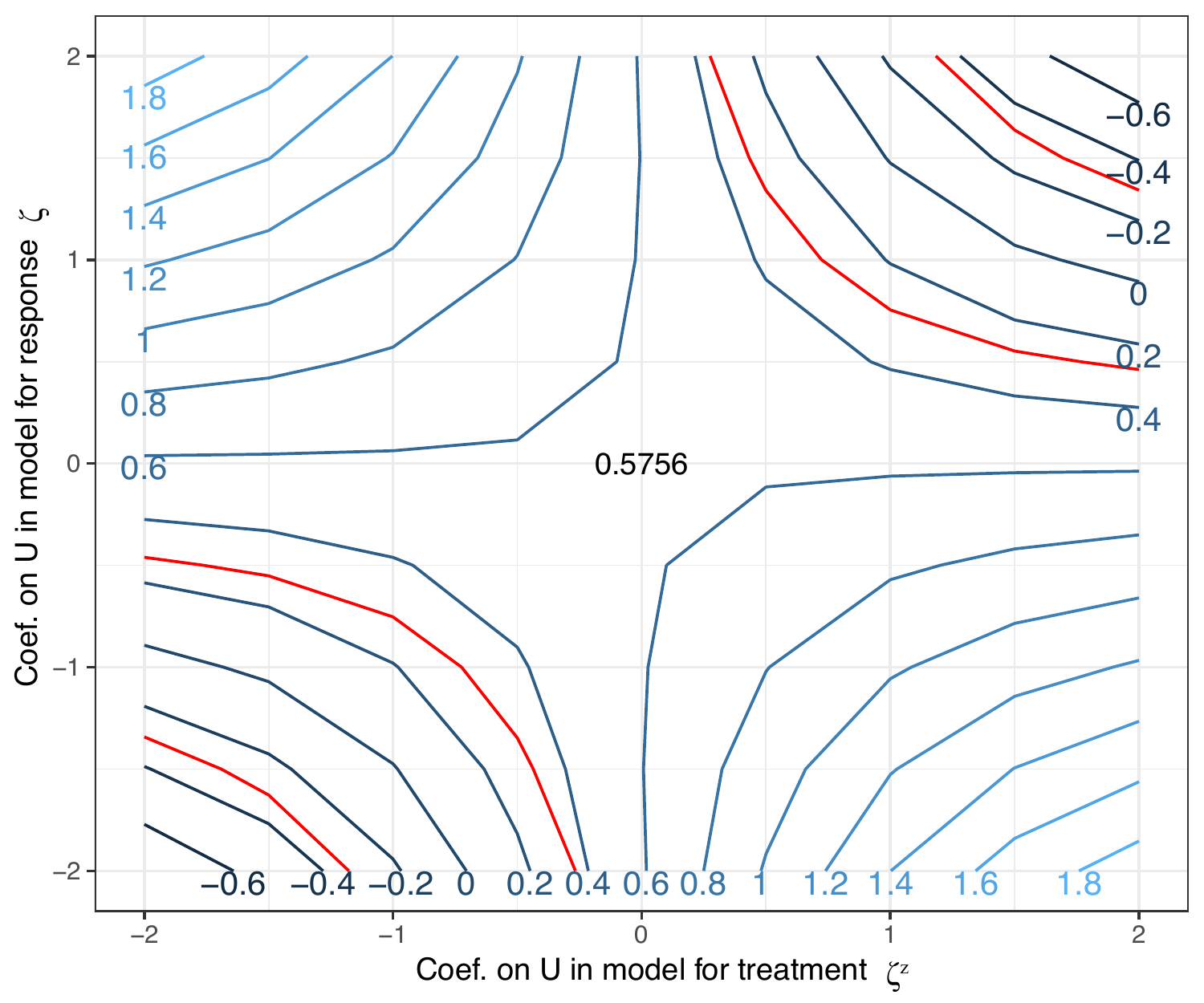}
		\label{fig:UCout1EM}
	}	
	\subfigure[IPW]{
		\includegraphics[width=0.45\textwidth]{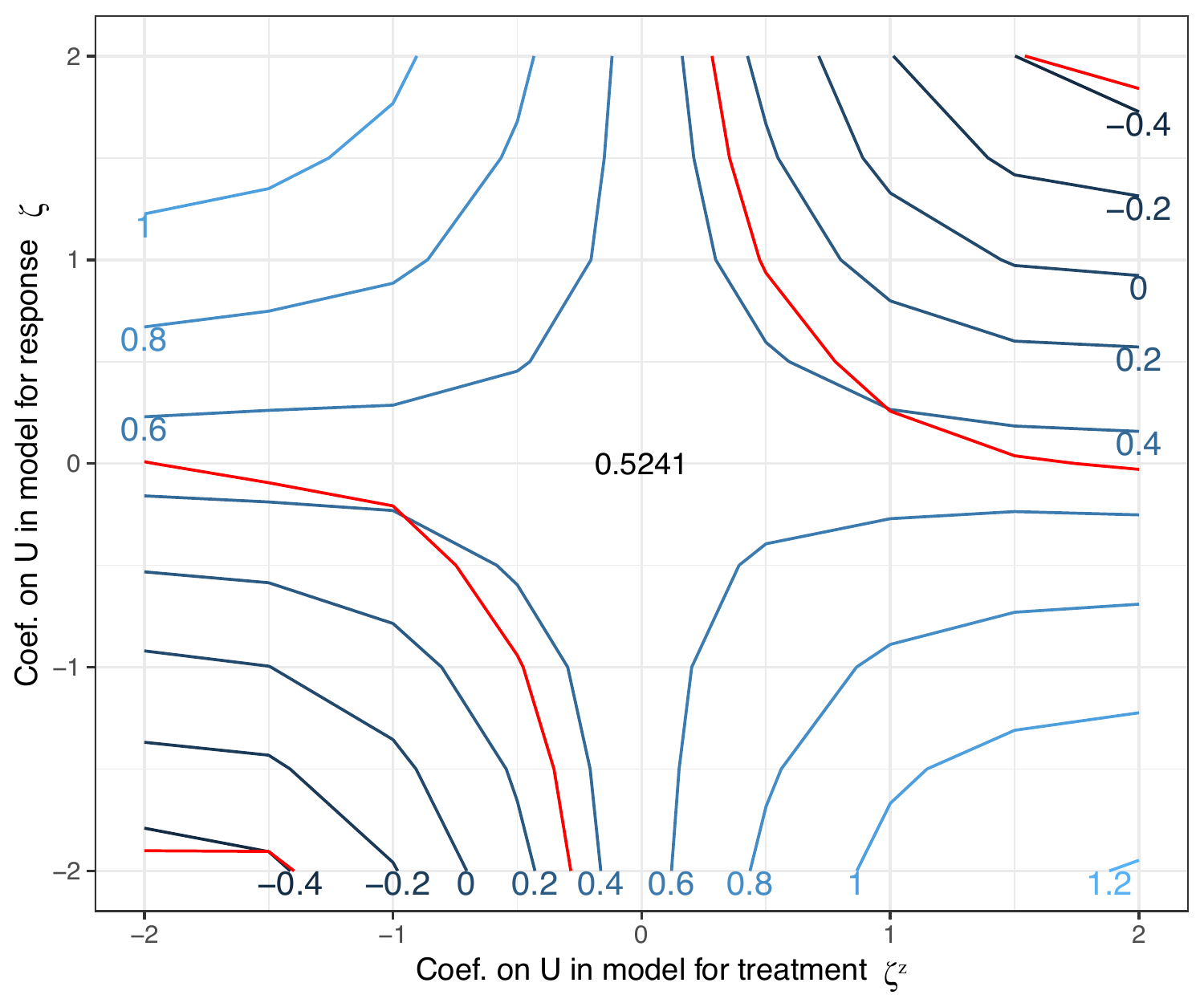}
		\label{fig:UCout1ipw}
	}		
	\caption{Sensitivity analysis results on the IBD for UC patients data for outcome clinical remission. In all plots, the blue contours show the sensitivity parameter values corresponding to the estimated treatment effect $\hat \tau$, and the red curves correspond to where the absolute value of the $t-$statistic $|t| = |\hat \tau / \hat \sigma_{\hat \tau}| = 1.96$. }
	\label{fig:UC}
\end{figure}

To be consistent with Lukin {\it et al.}\cite{dulai_UC}, here we consider a single covariate 
 $X_i = \Phi^{-1}(\ps_i)$ in our models, as this quantity is more likely to be normally distributed than $\ps_i$. We then assume that there is an unmeasured confounder $U \sim$ Bernoulli(0.5). 
 To determine the range for the sensitivity parameters, we take into consideration the observed association between a measured confounder and the treatment assignment or the outcome, in this case all less than one in absolute value in terms of log odds ratio or log hazard ratio. In addition, 
 a probit coefficient on a binary variable ($U$) is likely to lie in $[-2,2]$ in practice as suggested in Carnegie {\it et al.} \cite{carnegie2016assessing}.
 Similarly under the Cox PH model, the log hazard ratio of $\pm 2$ is very substantial for a binary variable. Therefore, we focused on $\zeta^z \in [-2,2]$ and $\zeta \in [-2,2]$. The EM and stochastic  EM algorithms were then applied as described in \secref{algo}. The estimates from the stochastic EM were obtained by averaging over $K=100$ estimates. The sensitivity analysis results are reported in \tabref{UCout1} and \figref{UC} panels (a) and (b).

In the models without unmeasured confounding ($\zeta^z = \zeta = 0$), the estimates were
 $\hat \beta^z$ (SE) = 1.1002 (0.0926), 
 $\hat \beta$ (SE) = -0.3250 (0.0994), and $\hat \tau$ (SE) = 0.5756 (0.1423), where `SE' stands for standard error. 
 We note that $\hat \beta^z$ would have been exactly one if, instead of `twang', probit regression had been used to fit the propensity score model.  
 In addition, the estimated treatment effect $\hat \tau$ here was obtained by regression adjustment, compared to the IPW estimate of Lukin.{\it et al.}\cite{dulai_UC} (see sensitivity analysis for IPW below also).
Nonetheless, the estimated treatment effects are qualitatively consistent:  Vedolizumab treated patients were more likely to achieve clinical remission compared to TNF-antagonist therapy, with hazard ratio (HR) = 1.7783 based on regression adjustment. 

 \figref{UC}(a) and (b) show that over a wide range of sensitivity parameters, the EM and the stochastic EM gave very similar results. 
 In the plots, the blue contours show the sensitivity parameter values corresponding to the estimated treatment effect $\hat \tau$, and the red curves correspond to where the absolute value of the $t-$statistic $|t| = |\hat \tau / \hat \sigma_{\hat \tau}| = 1.96$. Hence, any combination of $(\zeta^z, \zeta)$ in the region between two red curves in the upper right  or lower left quadrant leads to a non-significant estimated treatment effect at 0.05 level two-sided. Note that except for very small random fluctuation in the stochastic EM results, the contours and curves are symmetric about the origin $(\zeta^z, \zeta) = (0,0)$, where the estimated $\hat \tau = 0.5756$ is marked.
 
  From the plots we see that in order to drive the estimated treatment effect to zero, 
  $(\zeta^z, \zeta)$ will need to be close to (1.5, 1) or (1, 1.5), for example, 
  compared to $\hat \beta^z$ = 1.1002, $\hat \beta  = -0.3250$ in model \eqref{model:surv} above; such a very strong association between $U$ and the survival outcome seems unlikely. 
  We also noted earlier that the observed association between a measured confounder and  the outcome were all less than one in absolute value in terms of  log hazard ratio (the largest being just under 0.6 in absolute value).
   Similarly the red curves in \figref{UC}  show that in order to drive the estimate to be non-significant at 0.05 level, $(\zeta^z, \zeta)$ will need to be close to (1, 0.8), for example.

Finally, as  IPW with $\ps_i$  was the main statistical approach used  in Lukin {\it et al.}\cite{dulai_UC} to estimate the treatment effect, we also carried out sensitivity analysis for this approach. We implemented this by combining the stochastic EM with IPW as follows. At convergence of the algorithm we simulated $U_i$ and estimated the propensity score $\mathbb{P}(Z=1|X, U)$ by regressing $Z_i$ on $X_i = \Phi^{-1}(\ps_i)$  and the simulated $U_i$, $i=1, ...n$. Stabilized weights were obtained and further trimmed to be within (0.1, 10) if necessary.
The IPW approach was then applied. The final estimates were also obtained by averaging over $K=100$ estimates, with the corresponding standard errors obtained using \eqref{mi_var} where $ \hat \sigma^2_{\hat \tau_{k}} $ was the sandwich variance estimator 
following the IPW. 
The results are also reported in \tabref{UCout1} and \figref{UC}(c).
It is seen that unlike the regression adjustment results above, where the estimated treatment effect remained the same as long as $\zeta=0$, here instead the estimated treatment effect remained the same as long as 
$\zeta^z=0$. We also note much larger standard errors  when 
$|\zeta^z|$ is large, perhaps understandable as the treatment groups become more imbalanced. 
However, similar to the regression adjustment results above, in order to drive the estimated treatment effect to zero,   $(\zeta^z, \zeta)$ will need to be close to (1.5, 1) or (1, 1.5). 
  On the other hand,  the estimated treatment effect may become non-significant at 0.05 level if $(\zeta^z, \zeta) = (0.5, 1)$ or $(\zeta^z, \zeta) = (0.8, 0.5)$.

\subsection{Crohn's disease data}

Crohn's disease (CD) is another type of IBD that can cause inflammation along anywhere of the digestive tract. We are again interested in comparing the effectiveness between Vedolizumab and tumor necrosis factor (TNF)-antagonist therapy for CD patients.  
The data was collected between May 2014 and December 2017 from the North American based consortium registry. \cite{dulai2016real} A total of 1,242 patients were included (655 treated with Vedolizumab, 587 with TNF-antagonist therapy).  The primary interest is the treatment effect of Vedolizumab ($Z=1$) versus (TNF)-antagonist ($Z=0$) on clinical remission, which is defined as complete resolution of CD-related symptoms. In the Vedolizumab group, 196 patients had clinical remission and 9 had surgery, while in the TNF-antagonist group, 255 patients had clinical remission and 18 patients had surgery. Due to the presence of competing events, We applied our approach under the competing risks models to estimate the treatment effect of Vedolizumab. 

In Bohm {\it et al.}\cite{dulai_CD}, the propensity score for each subject $i$, denoted $\ps_i$, was calculated using the R package `twang'\cite{mccaffrey2004propensity} based on pre-treatment variables, including prior TNF-antagonist exposure and number of prior TNF-antagonists exposed, disease extent, history of fistulizing disease, prior bowel surgery, disease phentyope, clinical disease severity, CD related hospitalization within the preceding 1-year, baseline steroid dependency or refractoriness, concomitant steroid use, or concomitant immunomodulator use. It is unclear if all confounders have been included, though. Hence, sensitivity analysis is also necessary for this data.

\begin{figure}[!thpb]
	\centering	
	\subfigure[stochastic EM, $\zeta_2=0$]{
		\includegraphics[width=0.45\textwidth]{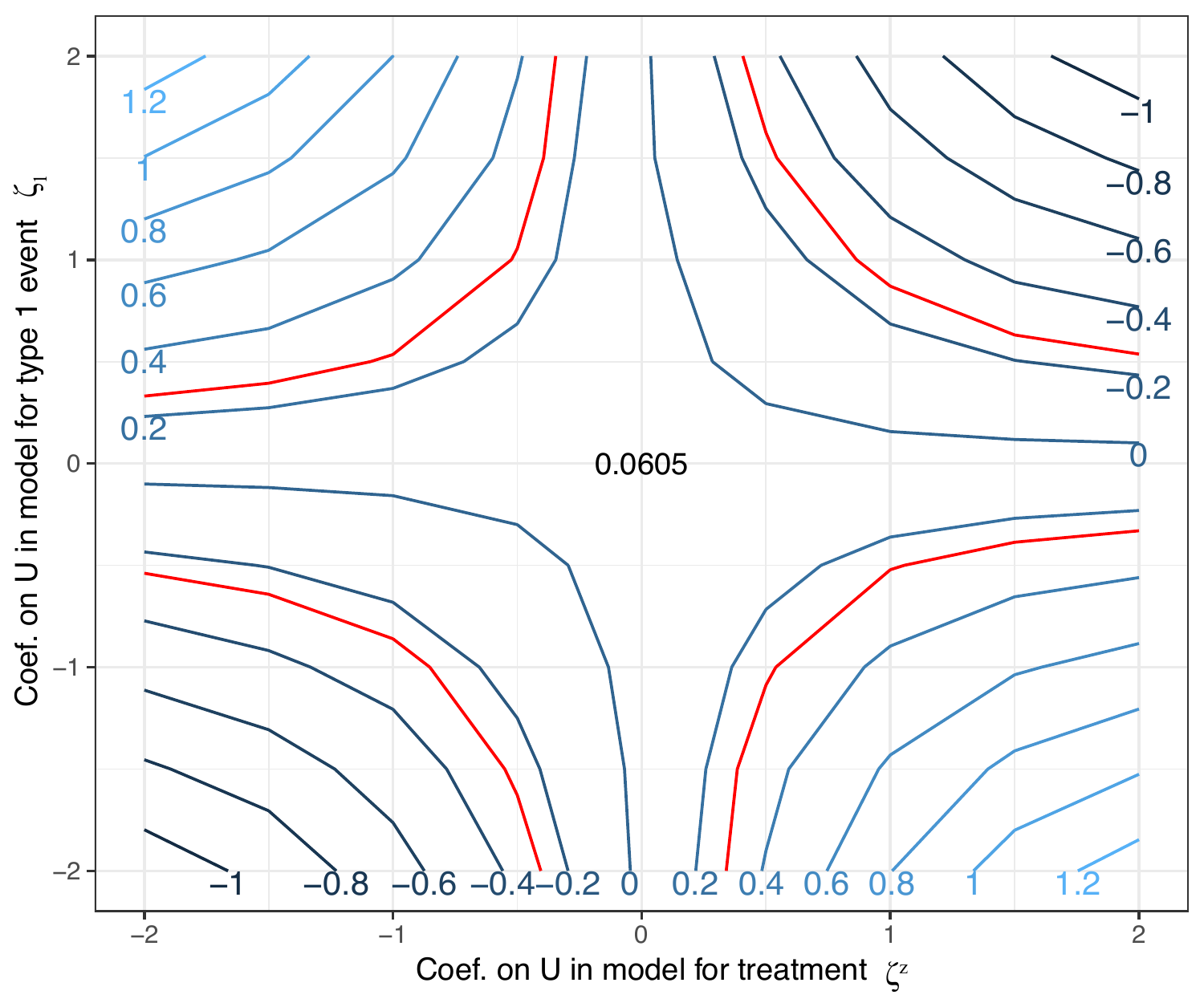}
		\label{fig:CDout10stoEM}
	}	
	\subfigure[EM, $\zeta_2=0$]{
		\includegraphics[width=0.45\textwidth]{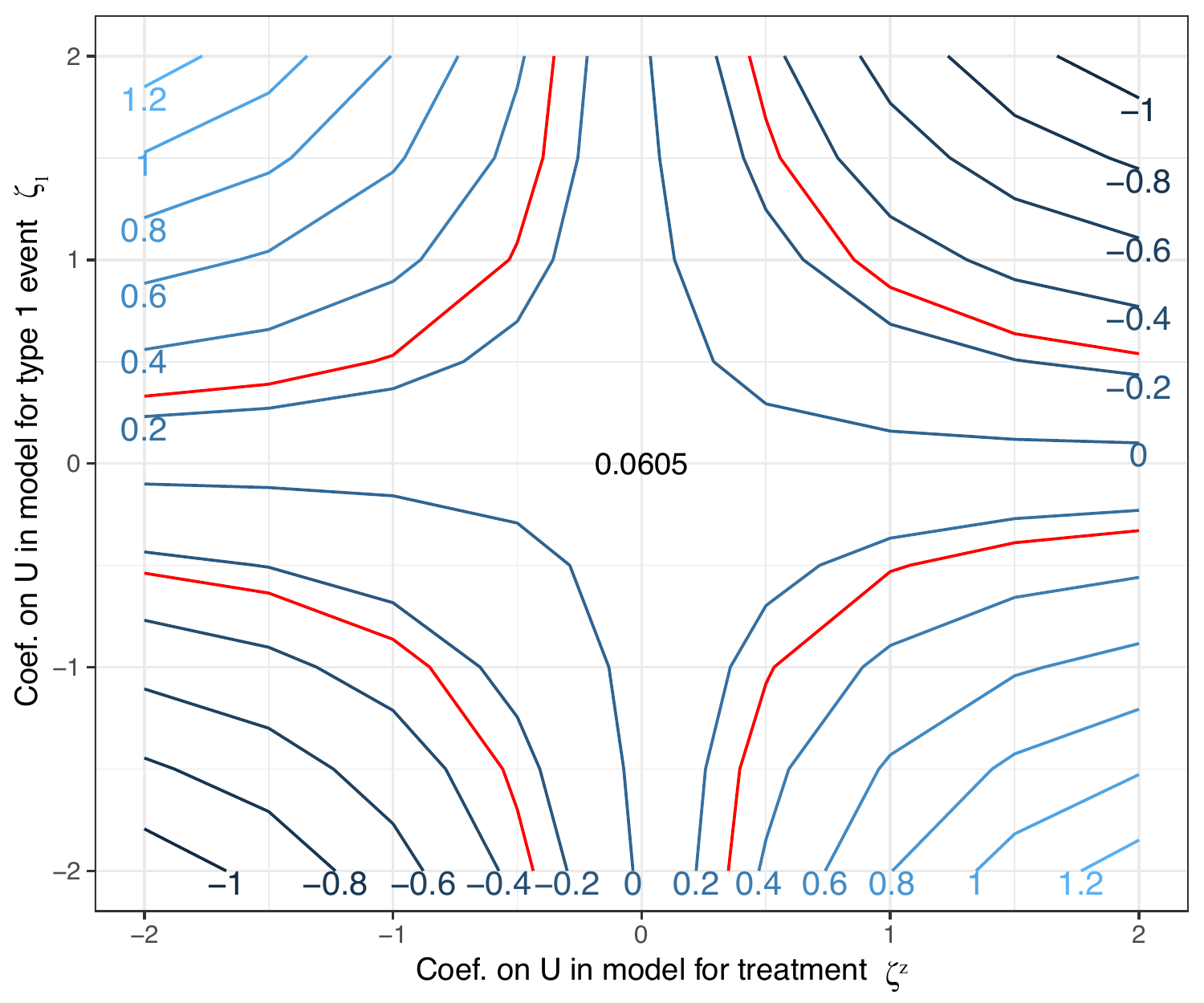}
		\label{fig:CDout10EM}
	}
	\subfigure[IPW, $\zeta_2=0$]{
		\includegraphics[width=0.45\textwidth]{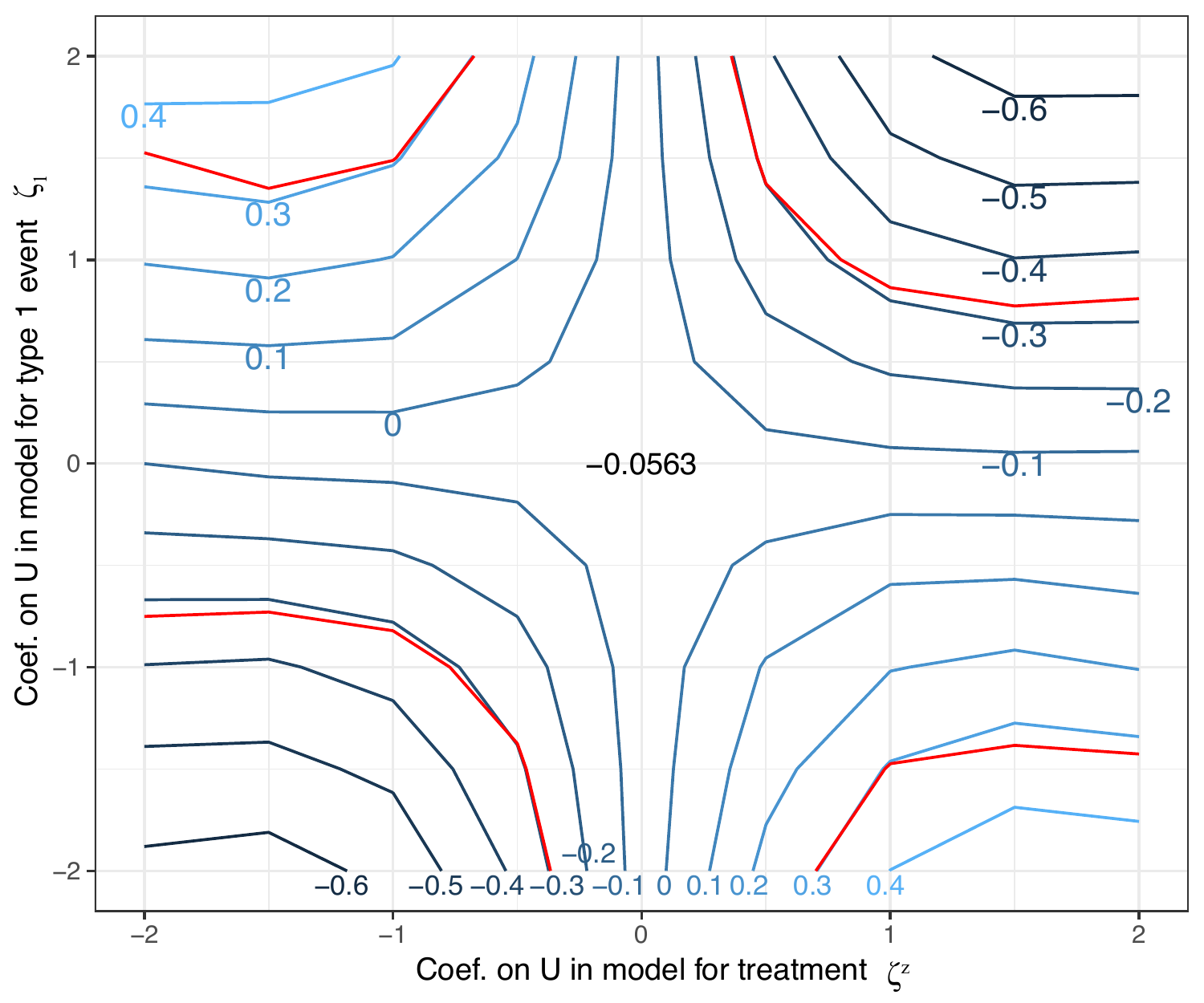}
		\label{fig:CDout10ipw}
	}	
	\caption{Sensitivity analysis results on the IBD for CD patients data for outcome clinical remission. In all plots, the blue contours show the values of $(\zeta^z, \zeta_1)$ corresponding to the estimated treatment effect $\hat \tau_1$, and the red curves correspond to where the absolute value of the $t-$statistic $|t| = |\hat \tau_i / \hat \sigma_{\hat \tau_1}| = 1.96$.}
	\label{fig:CD}
\end{figure}

To be consistent with Bohm {\it et al.}\cite{dulai_CD}, we consider a single covariate $X_i = \Phi^{-1}(\ps_i)$ in our models and assume an the unmeasured confounder $U \sim$ Bernoulli(0.5). The range for the sensitivity parameters is determined similarly as the UC data. We focus on $\zeta^z \in [-2,2]$ and $\zeta_1 \in [-2,2]$, and $\zeta_2 \in \{-2,0,2\}$. The EM and stochastic EM algorithms were then applied as described in \secref{algo}. The estimates from the stochastic EM were obtained by averaging over $K=100$ estimates. The sensitivity analysis results are reported in \tabref{CDout1_neg2},  \tabref{CDout1_0}, \tabref{CDout1_pos2} and \figref{CD} panels (a) and (b).

In the models without unmeasured confounding ($\zeta^z = \zeta_1 = \zeta_2 = 0$), the estimate of $\beta^z$ as defined in model \eqref{model:ps} is $\hat \beta^z$ (SE) = 1.0631 (0.0513), the estimates of $\beta_j$ $(j=1,2)$ as defined in model \eqref{model:comprisk} are  $\hat \beta_1$ (SE) = $-0.1664$ (0.0562) and $\hat \beta_2$ (SE) = $-$0.2601 (0.2401), and the estimates of $\tau_j$ $(j=1,2)$ are $\hat \tau_1$ (SE) = 0.0605 (0.1318)  and $\hat \tau_2$ (SE) = $-$0.0537 (0.5705). 
 If probit regression had been used to fit the propensity score model, $\hat \beta^z$ would have been exactly one.
 In addition, the estimated treatment effects $\hat \tau_j$ $(j=1,2)$  here were obtained by regression adjustment, compared to the IPW estimate of Bohm {\it et al.}\cite{dulai_CD}. Nonetheless, the estimated treatment effects are consistent: there was no significant difference in time to clinical remission between Vedolizumab and TNF-antagonist treated patients, with hazard ratio (HR) = 1.0624 based on regression adjustment. 

Note that by our algorithms, $\zeta_2$ affects $\hat \tau_1$ only through the conditional probability of $U$ as shown in \eqref{conU}. In this data, as the number of surgery is relatively small compared to the number of clinical remission, the effect of $\zeta_2$ on $\hat \tau_1$ is subtle (\tabref{CDout1_neg2},  \tabref{CDout1_0}, \tabref{CDout1_pos2}). This is, of course, not necessarily true when the number of the competing risk events is comparable to the number of events of interest.

 \figref{CD} (a) and (b) show that when $\zeta_2 = 0$, over a wide range of $(\zeta^z, \zeta_1)$, the EM and the stochastic EM gave similar results. In the plots, the blue contours show the values of $(\zeta^z, \zeta_1)$ corresponding to the estimated treatment effect $\hat \tau_1$, and the red curves correspond to where the absolute value of the $t-$statistic $|t| = |\hat \tau_i / \hat \sigma_{\hat \tau_1}| = 1.96$.  Hence, any combination of $(\zeta^z, \zeta_1)$ in the region surrounded by four red curves leads to a non-significant estimated treatment effect at level 0.05 two-sided. Except for very small random fluctuation in the stochastic EM results, the contours and curves are symmetric about the origin $(\zeta^z, \zeta_1) = (0,0)$, where the estimated $\hat \tau_1 = 0.0605$ is marked. In order to drive the estimated treatment effect to being significant, given $\zeta_2 = 0$, $(\zeta^z, \zeta_1)$ will need to be close to (1,1) or ($-$0.8, 0.8), for example, compared to  $\hat \beta^z$ = 1.0631 and $\hat \beta_1$ = $-$0.1664 above; such a strong association between $U$ and the outcome seems unlikely in practice.

Finally, as IPW with $\ps_i$  was the main statistical approach used  in Bohm {\it et al.}\cite{dulai_CD} to estimate the treatment effect, we also carried out sensitivity analysis for this approach by combing the stochastic EM with IPW as under the survival models. The final estimates were also obtained by averaging over $K=100$ estimates. The results are reported in \tabref{CDout1_neg2},  \tabref{CDout1_0}, \tabref{CDout1_pos2} and \figref{CD}(c). Similar to the regression adjustment results, in order to drive the estimated treatment effect to significant, given $\zeta_2 = 0$, $(\zeta^z, \zeta_1)$ will need to be close to (1, 1) or (-1, 1.5), which seems unlikely in practice. 

\section{Discussion} \label{sec:discussion}

In this paper we developed approaches to perform sensitivity analysis of the estimated treatment effect with regard to unobserved confounding in observational studies with survival or competing risks outcomes. 
The approaches we developed are based on models for survival or competing risks outcomes, which allow simulating the unobserved confounder given the observed data. The sensitivity parameters reflect the association between the unobserved confounder and the outcomes, as well as the association between the unobserved confounder and the treatment assignments. The interpretation of these sensitivity parameters is  straightforward, which leads to relative ease in choosing  plausible ranges  for them. 
Simulation studies show that both the EM and the stochastic EM algorithm are able to recover the true treatment effect if the correct sensitivity parameter values are used. The EM algorithm is clearly optimal in theory,\cite{nielsen2000stochastic} although the stochastic EM allows easy incorporation of IPW approaches for estimating treatment effects, which are commonly used in practice and as we have illustrated in our data analysis. 

For the distribution of the unobserved confounder we used binary 0, 1 with probability 0.5 each, which were recommended and used throughout the book by Rosenbaum.\cite{RosenbaumOS}
It is also possible to incorporate  normally distributed $U$, such as in Shen {\it et al.}\cite{shen2011sensitivity} 
and Xu {\it et al.},\cite{xu:etal19} in which case the probit link in model \eqref{model:ps} allows closed-form marginal propensity scores given $X$ after integrating out $U$. The $ \mathcal{Q}_1$ part of the EM algorithm would be similar to that under the proportional hazards mixed-effects model (PHMM) and Monte Carlo approximation would be needed in the E-steps.\cite{vaida2000proportional} 


Carnegie {\it et al.} \cite{carnegie2016assessing} discussed the advantages and disadvantages of using parametric versus nonparametric approaches in sensitivity analysis. Parametric approaches are typically needed in order to simulate the unobserved confounder; in survival analysis however, the outcome models are often semiparametric, allow flexibility in modeling in particular the nuisance parameters. 
On the other hand, nonparametric bounds might be considered under minimal assumptions in place of sensitivity analysis.\cite{richardson2014nonparametric}  However, such bounds can be very difficult to derive for complex outcomes like what we consider here in the presence of right censoring, which is unlike in Shen {\it et al.}\cite{shen2011sensitivity} where it is possible to derive these bounds for binary or continuous outcomes without censoring. Also evident in Shen {\it et al.}\cite{shen2011sensitivity} is that parametric settings are often needed in order to aid in the interpretation of the 
sensitivity parameters  in the corresponding nonparametric settings, and extensive simulations have to be conducted in order to determine sensible ranges for these sensitivity parameters.\cite{xu:etal19}

\vskip .2in
\centerline{\bf Acknowledgement}
This work was partially supported by National Institutes of Health, Grant UL1TR001442 of CTSA funding. 

\vskip .3in
\centerline{APPENDIX}

In the following we write out the components of $\ddot{l}$ and $s$ for competing risks with $j=1, ..., m$. For a single survival outcome without competing risks, we should simply take $m=1$ and the corresponding parameters are the same as without the subscript $j$.

The components of $s$ are: 
\begin{align}
    \frac{\partial l}{\partial \beta_j} &= \sum_{i=1}^n x_i \{\delta_{ij} - \Lambda_{j0}(t_i) \exp(\tau_j z_i + \beta_j'x_i + \zeta_j u_i )\}\\
    \frac{\partial l}{\partial \tau_j} &= \sum_{i=1}^n z_i \{\delta_{ij} - \Lambda_{j0}(t_i) \exp(\tau_j z_i + \beta_j'x_i + \zeta_j u_i )\} \\
    \frac{\partial l}{\partial \lambda_{j0}(t_i)} &= \frac{1}{\lambda_{j0}(t_i)} - \sum_{t_k \ge t_i} \exp(\tau_j z_k + \beta_j'x_k + \zeta_j u_k)\\
    \frac{\partial l}{\partial \beta^z} &= \sum_{i=1}^n \big \{z_i \frac{\phi(x_i' \beta^z + \zeta^z u_i)}{\Phi(x_i' \beta^z + \zeta^z u_i)} - (1-z_i) \frac{\phi(x_i' \beta^z + \zeta^z u_i)}{1-\Phi(x_i' \beta^z + \zeta^z u_i)}\big\}x_i
\end{align}
for $j=1,\cdots, m$. For the second derivatives,
\begin{align}
    \frac{\partial^2 l}{\partial \beta_j^2} &= - \sum_{i=1}^n x_i^{\otimes 2} \Lambda_{j0}(t_i) \exp( \tau_j z_i + \beta_j' x_i + \zeta_j u_i)\\
    \frac{\partial^2 l}{\partial \tau_j^2} &= - \sum_{i=1}^n z_i \Lambda_{j0}(t_i) \exp(\tau_j z_i + \beta_j'x_i + \zeta_j u_i) \\ 
    \frac{\partial^2 l}{\partial \lambda_{j0}(t_i)^2} &= -\frac{1}{\lambda_{j0}(t_i)^2}\\
    \frac{\partial^2 l}{\partial \beta_j \partial \tau_j} &= - \sum_{i=1}^n z_i x_i \Lambda_{j0}(t_i) \exp(\tau_j z_i + \beta_j' x_i + \zeta_j u_i)\\
    \frac{\partial^2 l}{\partial \beta_j \partial \lambda_{j0}(t_i)} &= - \sum_{t_k \ge t_i} x_k \exp(\tau_j z_k + \beta_j'x_k + \zeta_j u_k)\\
    \frac{\partial^2 l}{\partial \tau_j \partial \lambda_{j0}(t_i)} &= - \sum_{t_k \ge t_i} z_k \exp(\tau_j z_k + \beta_j'x_k + \zeta_j u_k) \\
    \frac{\partial^2 l}{\partial \beta^{z2}} &= - \sum_{i=1}^n \phi(x_i' \beta^z + \zeta^z u_i) \big \{z_i \frac{\phi(x_i' \beta^z + \zeta^z u_i) + (x_i' \beta^z + \zeta^z u_i) \Phi(x_i' \beta^z + \zeta^z u_i) }{\Phi(x_i' \beta^z + \zeta^z u_i)^2} \nonumber \\
    & + (1-z_i) \frac{\phi(x_i' \beta^z + \zeta^z u_i) - (x_i' \beta^z + \zeta^z u_i) (1-\Phi(x_i' \beta^z + \zeta^z u_i)) }{(1-\Phi(x_i' \beta^z + \zeta^z u_i))^2}\big\}x_i^{\otimes 2}
\end{align}
where $a^{\otimes 2} = a a'$ for a vector $a$, $\phi$ is the probability density function (pdf) of the standard normal distribution, and all other off-diagonal elements are zeros. The computation of the first term in \eqref{info} is similar to the computation in the E-step for different functions $h(u_i)$. To calculate the second term in \eqref{info}, we sample $U$ from Bernoulli$(\tilde \pi)$ for 1,000 times after convergence of the EM, and  take the average of $s(\theta;\mathbf{y},\mathbf{z},\mathbf{u})s(\theta;\mathbf{y},\mathbf{z},\mathbf{u})'$ over the sampled $U$'s.

\bibliographystyle{WileyNJD-AMA}
\bibliography{propensity}

\newpage

\begin{figure}[!thpb]
\centering
\includegraphics[width=1.0\textwidth]{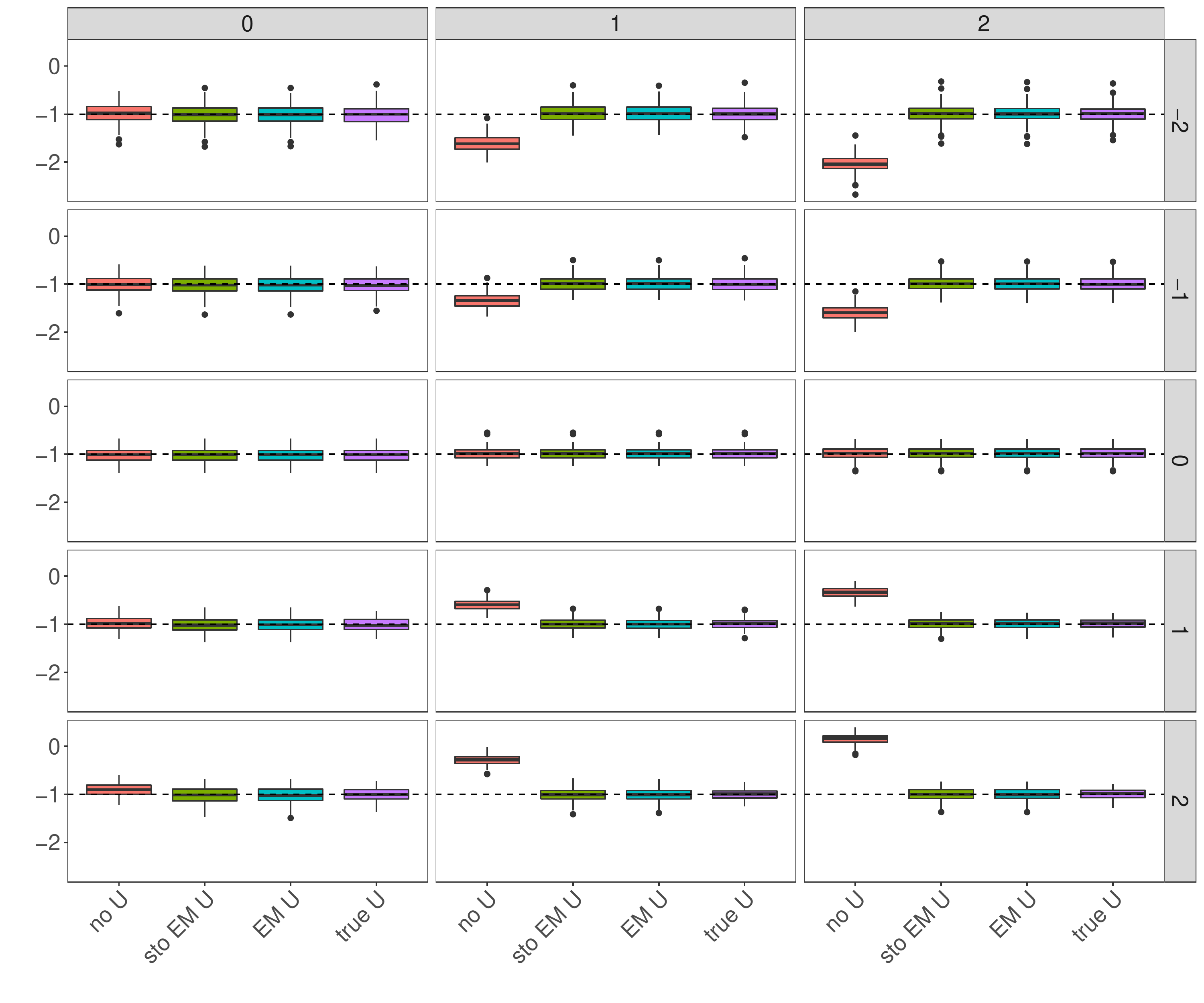}
\caption{Distributions of the estimated treatment effect on type 2 failures for the simulated competing risks data. $\zeta^z \in \{0,1,2\}$ on the horizontal label and $\zeta_1 = \zeta_2 \in \{-2,-1, 0,1,2\}$ on the vertical label. Each boxplot displays $\hat \tau_2$ from 200 simulations. }
\label{fig:tau2}
\end{figure}

\newpage

\begin{table}[!thpb]
\caption{Estimated treatment effect  (standard deviation) for the simulated survival data with $\tau=1$}
\begin{center}
\begin{tabular}{c l c c c}
\hline
\hline 
& Method & $\zeta^z = 0$ & $\zeta^z = 1$ & $\zeta^z = 2$ \\
\hline
\multirow{4}{4em}{$\zeta=-2$} & True $U$ & 1.0171 (0.1244) & 1.0108 (0.1171) & 1.0033 (0.1166) \\
 & EM & 1.0216 (0.1496) & 1.0101 (0.1469) & 1.0015 (0.1297) \\
 & Sto EM & 1.0257 (0.1502) & 1.0132 (0.1468) & 1.0066 (0.1296) \\
 & No $U$ & 0.7873 (0.1219) & 0.1512 (0.1257) & -0.2052 (0.1141) \\ [0.8em]
\multirow{4}{4em}{$\zeta=-1$} & True $U$ & 1.0206 (0.1015) & 1.0144 (0.1067) & 1.0129 (0.1071) \\
 & EM & 1.0203 (0.1153) & 1.0121 (0.1173) & 1.0104 (0.1103) \\
 & Sto EM & 1.0220 (0.1157) & 1.0125 (0.1172) & 1.0118 (0.1105) \\
 & No $U$ & 0.9310 (0.1068) & 0.5664 (0.1109) & 0.3524 (0.1068) \\ [0.8em]
\multirow{4}{4em}{$\zeta=0$} & True $U$ & 1.0159 (0.0868) & 1.0124 (0.0996) & 1.0095 (0.1035) \\
 & EM & 1.0159 (0.0868) & 1.0124 (0.0996) & 1.0095 (0.1035) \\
 & Sto EM & 1.0159 (0.0868) & 1.0124 (0.0996) & 1.0095 (0.1035) \\
 & No $U$ & 1.0159 (0.0868) & 1.0124 (0.0996) & 1.0095 (0.1035) \\ [0.8em]
\multirow{4}{4em}{$\zeta=1$} & True $U$ & 1.0148 (0.0797) & 1.0134 (0.0896) & 1.0110 (0.1004) \\
 & EM & 1.0188 (0.0891) & 1.0167 (0.0977) & 1.0139 (0.1072) \\
 & Sto EM & 1.0195 (0.0894) & 1.0183 (0.0977) & 1.0164 (0.1068) \\
 & No $U$ & 0.9059 (0.0802) & 1.2601 (0.0878) & 1.4993 (0.0971) \\ [0.8em]
\multirow{4}{4em}{$\zeta=2$} & True $U$ & 1.0133 (0.0768) & 1.0164 (0.0875) & 1.0154 (0.1031) \\
 & EM & 1.0226 (0.1047) & 1.0260 (0.1122) & 1.0263 (0.1218) \\
 & Sto EM & 1.0225 (0.1052) & 1.0271 (0.1127) & 1.0303 (0.1228) \\
 & No $U$ & 0.6946 (0.0783) & 1.2618 (0.0835) & 1.6734 (0.0942) \\
\hline
\hline 
\end{tabular}
\end{center}
\label{tab:tau}
\end{table}

\begin{table}[!thpb]
\caption{Treatment effect estimate (standard deviation) on type 1 failures for the simulated competing risks data with $\tau_1 = 1$}
\begin{center}
\begin{tabular}{c l c c c}
\hline
\hline 
& method & $\zeta^z = 0$ & $\zeta^z = 1$ & $\zeta^z = 2$ \\
\hline
\multirow{4}{6em}{$\zeta_1 = \zeta_2 = -2$} & True $U$ & 1.0150 (0.1428) & 1.0293 (0.1603) & 1.0305 (0.1644) \\
 & EM & 1.0154 (0.1767) & 1.0357 (0.1801) & 1.0354 (0.1831) \\
 & StoEM & 1.0180 (0.1788) & 1.0390 (0.1825) & 1.0428 (0.1844) \\
 & No $U$ & 0.9312 (0.1583) & 0.3192 (0.1670) & -0.0215 (0.1678) \\ [0.8em]
\multirow{4}{6em}{$\zeta_1 = \zeta_2 = -1$} & True $U$ & 1.0141 (0.1327) & 1.0185 (0.1542) & 1.0269 (0.1613) \\
 & EM & 1.0141 (0.1390) & 1.0186 (0.1585) & 1.0280 (0.1677) \\
 & StoEM & 1.0150 (0.1388) & 1.0191 (0.1584) & 1.0304 (0.1675) \\
 & No $U$ & 0.9817 (0.1339) & 0.6243 (0.1536) & 0.4260 (0.1635) \\ [0.8em]
\multirow{4}{6em}{$\zeta_1 = \zeta_2 = 0$} & True $U$ & 1.0153 (0.1212) & 1.0258 (0.1329) & 1.0317 (0.1593) \\
 & EM & 1.0153 (0.1212) & 1.0258 (0.1329) & 1.0317 (0.1593) \\
 & StoEM & 1.0153 (0.1212) & 1.0258 (0.1329) & 1.0317 (0.1593) \\
 & No $U$ & 1.0153 (0.1212) & 1.0258 (0.1329) & 1.0317 (0.1593) \\ [0.8em]
\multirow{4}{6em}{$\zeta_1 = \zeta_2 = 1$} & True $U$ & 1.0078 (0.1115) & 1.0276 (0.1259) & 1.0348 (0.1524) \\
 & EM & 1.0088 (0.1226) & 1.0251 (0.1415) & 1.0320 (0.1549) \\
 & StoEM & 1.0095 (0.1230) & 1.0272 (0.1422) & 1.0346 (0.1549) \\
 & No $U$ & 0.9745 (0.1130) & 1.3518 (0.1319) & 1.5750 (0.1470) \\ [0.8em]
\multirow{4}{6em}{$\zeta_1 = \zeta_2 = 2$} & True $U$ & 1.0064 (0.1094) & 1.0263 (0.1238) & 1.0345 (0.1570) \\
 & EM & 1.0022 (0.1370) & 1.0174 (0.1600) & 1.0266 (0.1645) \\
 & StoEM & 1.0031 (0.1369) & 1.0230 (0.1615) & 1.0345 (0.1652) \\
 & No $U$ & 0.9238 (0.1141) & 1.5064 (0.1300) & 1.8424 (0.1411) \\
\hline
\hline 
\end{tabular}
\end{center}
\label{tab:tau1}
\end{table}

\begin{table}[!thpb]
\caption{Treatment effect estimate (standard deviation) on type 2 failures for the simulated competing risks data with $\tau_2 = -1$}
\begin{center}
\begin{tabular}{c l c c c}
\hline
\hline 
& method & $\zeta^z = 0$ & $\zeta^z = 1$ & $\zeta^z = 2$ \\
\hline
\multirow{4}{6em}{$\zeta_1 = \zeta_2 = -2$} & true $U$ & -1.0170 (0.1974) & -0.9969 (0.1829) & -0.9994 (0.1707) \\
 & EM & -1.0170 (0.2143) & -0.9892 (0.1837) & -0.9937 (0.1826) \\
 & stoEM & -1.0176 (0.2163) & -0.9892 (0.1845) & -0.9911 (0.1822) \\
 & no $U$ & -0.9893 (0.2002) & -1.6139 (0.1713) & -2.0409 (0.1709) \\ [0.8em]
\multirow{4}{6em}{$\zeta_1 = \zeta_2 = -1$} & true $U$ & -1.0236 (0.1702) & -0.9996 (0.1484) & -0.9964 (0.1469) \\
 & EM & -1.0231 (0.1772) & -0.9945 (0.1518) & -0.9940 (0.1517) \\
 & stoEM & -1.0233 (0.1774) & -0.9951 (0.1518) & -0.9936 (0.1515) \\
 & no $U$ & -1.0109 (0.1728) & -1.3505 (0.1482) & -1.5975 (0.1485) \\ [0.8em]
\multirow{4}{6em}{$\zeta_1 = \zeta_2 = 0$} & true $U$ & -1.0152 (0.1472) & -0.9851 (0.1189) & -0.9797 (0.1238) \\
 & EM & -1.0152 (0.1472) & -0.9851 (0.1189) & -0.9797 (0.1238) \\
 & stoEM & -1.0152 (0.1472) & -0.9851 (0.1189) & -0.9797 (0.1238) \\
 & no $U$ & -1.0152 (0.1472) & -0.9851 (0.1189) & -0.9797 (0.1238) \\ [0.8em]
\multirow{4}{6em}{$\zeta_1 = \zeta_2 = 1$} & true $U$ & -1.0088 (0.1316) & -0.9886 (0.1019) & -0.9836 (0.1097) \\
 & EM & -1.0108 (0.1423) & -0.9950 (0.1109) & -0.9887 (0.1115) \\
 & stoEM & -1.0110 (0.1434) & -0.9953 (0.1117) & -0.9883 (0.1114) \\
 & no $U$ & -0.9776 (0.1328) & -0.5956 (0.1039) & -0.3335 (0.1074) \\ [0.8em]
\multirow{4}{6em}{$\zeta_1 = \zeta_2 = 2$} & true $U$ & -1.0069 (0.1281) & -0.9950 (0.0976) & -0.9910 (0.1069) \\
 & EM & -1.0180 (0.1647) & -1.0093 (0.1322) & -1.0029 (0.1275) \\
 & stoEM & -1.0184 (0.1662) & -1.0070 (0.1338) & -0.9983 (0.1282) \\
 & no $U$ & -0.9075 (0.1313) & -0.2871 (0.1038) & 0.1522 (0.1050) \\
\hline
\hline 
\end{tabular}
\end{center}
\label{tab:tau2}
\end{table}

\begin{table}[!thpb]
	\caption{Treatment effect estimate (standard deviation) on type 1 failures for the simulated competing risks data with $\tau_1 = 1$ and $\zeta_1 = 1$ fixed.}
	\begin{center}
		\begin{tabular}{c l c c c}
		\hline
		\hline 
		& method & $\zeta^z = 0$ & $\zeta^z = 1$ & $\zeta^z = 2$ \\
		\hline
		\multirow{4}{4em}{$\zeta_2 = -2$} & true $U$ & 0.9971 (0.1011) & 1.0187 (0.1163) & 1.0271 (0.1505) \\
		& EM & 0.9963 (0.1093) & 1.0174 (0.1320) & 1.0263 (0.1545) \\
		& stoEM & 0.9971 (0.1101) & 1.0189 (0.1320) & 1.0282 (0.1549) \\
		& no $U$ & 0.8893 (0.0998) & 1.2872 (0.1214) & 1.5816 (0.1468) \\ [0.8em]
		\multirow{4}{4em}{$\zeta_2 = -1$} & true $U$ & 1.0020 (0.1008) & 1.0212 (0.1227) & 1.0292 (0.1508) \\
		& EM & 1.0028 (0.1080) & 1.0203 (0.1369) & 1.0280 (0.1544) \\
		& stoEM & 1.0034 (0.1091) & 1.0214 (0.1368) & 1.0299 (0.1548) \\
		& no $U$ & 0.9023 (0.0992) & 1.2980 (0.1264) & 1.5844 (0.1465) \\ [0.8em]
		\multirow{4}{4em}{$\zeta_2 = 0$} & true $U$ & 1.0033 (0.0966) & 1.0271 (0.1183) & 1.0326 (0.1481) \\
		& EM & 1.0040 (0.1045) & 1.0258 (0.1338) & 1.0305 (0.1507) \\
		& stoEM & 1.0050 (0.1045) & 1.0275 (0.1338) & 1.0325 (0.1510) \\
		& no $U$ & 0.9243 (0.0965) & 1.3220 (0.1244) & 1.5863 (0.1431) \\ [0.8em]
		\multirow{4}{4em}{$\zeta_2 = 1$} & true $U$ & 1.0078 (0.1115) & 1.0276 (0.1259) & 1.0348 (0.1524) \\
		& EM & 1.0088 (0.1226) & 1.0251 (0.1415) & 1.0320 (0.1549) \\
		& stoEM & 1.0095 (0.1230) & 1.0272 (0.1422) & 1.0346 (0.1549) \\
		& no U & 0.9745 (0.1130) & 1.3518 (0.1319) & 1.5750 (0.1470) \\ [0.8em]
		\multirow{4}{4em}{$\zeta_2 = 2$} & true $U$ & 1.0148 (0.1231) & 1.0245 (0.1402) & 1.0301 (0.1592) \\
		& EM & 1.0135 (0.1353) & 1.0195 (0.1506) & 1.0255 (0.1609) \\
		& stoEM & 1.0146 (0.1356) & 1.0223 (0.1511) & 1.0295 (0.1615) \\
		& no $U$ & 1.0452 (0.1271) & 1.3591 (0.1415) & 1.5158 (0.1520) \\
		\hline
		\hline 
		\end{tabular}
	\end{center}
	\label{tab:tau1_1}
\end{table}

\begin{table}[!thpb]
	\caption{Treatment effect estimate (standard deviation) on type 2 failures for the simulated competing risks data with $\tau_2 = -1$ and $\zeta_1 = 1$ fixed.}
	\begin{center}
		\begin{tabular}{c l c c c}
		\hline
		\hline 
		& method & $\zeta^z = 0$ & $\zeta^z = 1$ & $\zeta^z = 2$ \\
		\hline
		\multirow{4}{4em}{$\zeta_2 = -2$} & true $U$ & -1.0113 (0.1926) & -1.0100 (0.1818) & -1.0107 (0.1747) \\
		& EM & -1.0118 (0.2107) & -0.9985 (0.1886) & -1.0031 (0.1832) \\
		& stoEM & -1.0142 (0.2111) & -1.0001 (0.1898) & -1.0002 (0.1831) \\
		& no $U$ & -0.8715 (0.1996) & -1.4541 (0.1786) & -1.8513 (0.1766) \\ [0.8em]
		\multirow{4}{4em}{$\zeta_2 = -1$} & true $U$ & -1.0224 (0.1803) & -1.0066 (0.1556) & -1.0082 (0.1537) \\
		& EM & -1.0222 (0.1887) & -0.9995 (0.1546) & -1.0038 (0.1567) \\
		& stoEM & -1.0227 (0.1888) & -1.0001 (0.1549) & -1.0035 (0.1568) \\
		& no $U$ & -0.9510 (0.1841) & -1.2829 (0.1511) & -1.5228 (0.1547) \\ [0.8em]
		\multirow{4}{4em}{$\zeta_2 = 0$} & true $U$ & -1.0170 (0.1693) & -1.0010 (0.1383) & -0.9957 (0.1329) \\
		& EM & -1.0170 (0.1693) & -1.0010 (0.1383) & -0.9957 (0.1329) \\
		& stoEM & -1.0170 (0.1693) & -1.0010 (0.1383) & -0.9957 (0.1329) \\
		& no $U$ & -1.0170 (0.1693) & -1.0010 (0.1383) & -0.9957 (0.1329) \\ [0.8em]
		\multirow{4}{4em}{$\zeta_2 = 1$} & true $U$ & -1.0088 (0.1316) & -0.9886 (0.1019) & -0.9836 (0.1097) \\
		& EM & -1.0108 (0.1423) & -0.9950 (0.1109) & -0.9887 (0.1115) \\
		& stoEM & -1.0110 (0.1434) & -0.9953 (0.1117) & -0.9883 (0.1114) \\
		& no $U$ & -0.9776 (0.1328) & -0.5956 (0.1039) & -0.3335 (0.1074) \\ [0.8em]
		\multirow{4}{4em}{$\zeta_2 = 2$} & true $U$ & -1.0054 (0.1153) & -0.9918 (0.0933) & -0.9876 (0.0969) \\
		& EM & -1.0143 (0.1513) & -1.0037 (0.1272) & -0.9976 (0.1175) \\
		& stoEM & -1.0139 (0.1522) & -1.0026 (0.1276) & -0.9927 (0.1174) \\
		& no $U$ & -0.7921 (0.1136) & -0.1505 (0.0981) & 0.3065 (0.0977) \\
		\hline
		\hline 
		\end{tabular}
	\end{center}
	\label{tab:tau2_1}
\end{table}

\begin{landscape}
\begin{table}[!thpb]
\small
\caption{Sensitivity analysis results on the IBD for UC patients data for outcome clinical remission}
\begin{center}
\begin{tabular}{clccccccc}
\hline
\hline 
                  $\zeta$ & method & $\zeta^z = 0$   & $\zeta^z = 0.5$ & $\zeta^z = 1$ & $\zeta^z = 1.5$  & $\zeta^z = 2$    \\
\hline                            
		\multirow{3}{2em}{-2} & EM & 0.5833 (0.1825) & 1.0176 (0.1780) & 1.3987 (0.1731) & 1.6922 (0.1690) & 1.9005 (0.1677) \\
		& stoEM & 0.5948 (0.1718) & 1.0238 (0.1702) & 1.4000 (0.1637) & 1.7116 (0.1658) & 1.9064 (0.1629) \\
		& IPW & 0.5241 (0.1563) & 0.8406 (0.1691) & 1.0589 (0.1866) & 1.1615 (0.2019) & 1.2121 (0.2125) \\ [0.8em]
		\multirow{3}{2em}{-1.5} & EM & 0.5954 (0.1640) & 0.9119 (0.1626) & 1.1886 (0.1616) & 1.4033 (0.1582) & 1.5570 (0.1560) \\
		& stoEM & 0.5866 (0.1601) & 0.9148 (0.1605) & 1.1917 (0.1570) & 1.4016 (0.1558) & 1.5672 (0.1535) \\
		& IPW & 0.5241 (0.1563) & 0.7763 (0.1703) & 0.9701 (0.1981) & 1.0716 (0.2072) & 1.0961 (0.2183) \\  [0.8em]
		\multirow{3}{2em}{-1} & EM & 0.5895 (0.1523) & 0.7950 (0.1520) & 0.9759 (0.1506) & 1.1185 (0.1495) & 1.2217 (0.1485) \\
		& stoEM & 0.5894 (0.1532) & 0.7964 (0.1507) & 0.9691 (0.1488) & 1.1197 (0.1491) & 1.2212 (0.1473) \\
		& IPW & 0.5241 (0.1563) & 0.7126 (0.1704) & 0.8327 (0.1985) & 0.8835 (0.2207) & 0.9221 (0.2387) \\  [0.8em]
		\multirow{3}{2em}{-0.5} & EM & 0.5798 (0.1449) & 0.6811 (0.1448) & 0.7711 (0.1445) & 0.8431 (0.1442) & 0.8956 (0.1440) \\
		& stoEM & 0.5799 (0.1450) & 0.6782 (0.1451) & 0.7732 (0.1443) & 0.8416 (0.1444) & 0.8987 (0.1439) \\
		& IPW & 0.5241 (0.1563) & 0.6210 (0.1693) & 0.6862 (0.2079) & 0.7287 (0.2315) & 0.7246 (0.2309) \\  [0.8em]
		\multirow{3}{2em}{0} & EM & 0.5756 (0.1424) & 0.5756 (0.1424) & 0.5756 (0.1424) & 0.5756 (0.1424) & 0.5756 (0.1424) \\
		& stoEM & 0.5756 (0.1423) & 0.5756 (0.1423) & 0.5756 (0.1423) & 0.5756 (0.1423) & 0.5756 (0.1423) \\
		& IPW & 0.5241 (0.1563) & 0.5216 (0.1700) & 0.4979 (0.2088) & 0.4847 (0.2391) & 0.4730 (0.2506) \\  [0.8em]
		\multirow{3}{2em}{0.5} & EM & 0.5798 (0.1449) & 0.4777 (0.1451) & 0.3852 (0.1445) & 0.3105 (0.1441) & 0.2561 (0.1436) \\
		& stoEM & 0.5829 (0.1447) & 0.4774 (0.1443) & 0.3875 (0.1448) & 0.3135 (0.1442) & 0.2573 (0.1438) \\
		& IPW & 0.5241 (0.1563) & 0.4196 (0.1701) & 0.3132 (0.2012) & 0.2540 (0.2243) & 0.2408 (0.2470) \\  [0.8em]
		\multirow{3}{2em}{1} & EM & 0.5895 (0.1524) & 0.3811 (0.1516) & 0.1923 (0.1508) & 0.0400 (0.1490) & -0.0701 (0.1477) \\
		& stoEM & 0.5888 (0.1493) & 0.3850 (0.1502) & 0.1844 (0.1498) & 0.0442 (0.1506) & -0.0688 (0.1468) \\
		& IPW & 0.5241 (0.1563) & 0.3156 (0.1675) & 0.1231 (0.1971) & -0.0152 (0.2264) & -0.0443 (0.2486) \\  [0.8em]
		\multirow{3}{2em}{1.5} & EM & 0.5954 (0.1641) & 0.2767 (0.1629) & -0.0105 (0.1581) & -0.2419 (0.1587) & -0.4084 (0.1541) \\
		& stoEM & 0.5901 (0.1618) & 0.2699 (0.1551) & -1e-04 (0.1574) & -0.2469 (0.1549) & -0.4040 (0.1528) \\
		& IPW & 0.5241 (0.1563) & 0.2280 (0.1652) & -0.0649 (0.1924) & -0.2371 (0.2213) & -0.2932 (0.2340) \\  [0.8em]
		\multirow{3}{2em}{2} & EM & 0.5833 (0.1784) & 0.1560 (0.1756) & -0.2250 (0.1743) & -0.5365 (0.1720) & -0.7611 (0.1615) \\
		& stoEM & 0.5927 (0.1677) & 0.1459 (0.1631) & -0.2268 (0.1685) & -0.5524 (0.1650) & -0.7519 (0.1614) \\
		& IPW & 0.5241 (0.1563) & 0.1434 (0.1610) & -0.1950 (0.1855) & -0.3996 (0.2070) & -0.5294 (0.2309) \\
\hline
\hline 
\end{tabular}
\end{center}
\label{tab:UCout1}
\end{table}
\end{landscape}

\begin{landscape}
\begin{table}[!thpb]
\small
\caption{Sensitivity analysis results on the IBD for CD patients data for outcome clinical remission with $\zeta_2 = -2$}
\begin{center}
\begin{tabular}{clccccccc}
\hline
\hline 
                  $\zeta_1$ & method & $\zeta^z = 0$   & $\zeta^z = 0.5$ & $\zeta^z = 1$ & $\zeta^z = 1.5$  & $\zeta^z = 2$    \\
\hline  
		\multirow{3}{2em}{-2} & EM & 0.0229 (0.1535) & 0.4237 (0.1546) & 0.8001 (0.1588) & 1.0966 (0.1551) & 1.2968 (0.1509) \\
		& stoEM & 0.0336 (0.1486) & 0.4195 (0.1485) & 0.7958 (0.1509) & 1.1031 (0.1490) & 1.2963 (0.1480) \\
		& IPW & -0.0563 (0.1496) & 0.2309 (0.1501) & 0.4044 (0.154) & 0.4651 (0.1586) & 0.4516 (0.1642) \\ [0.8em]
		\multirow{3}{2em}{-1.5} & EM & 0.0403 (0.1471) & 0.3505 (0.1461) & 0.6294 (0.1463) & 0.8422 (0.1438) & 0.9854 (0.1433) \\
		& stoEM & 0.0394 (0.1421) & 0.3488 (0.1441) & 0.6279 (0.1433) & 0.8475 (0.1422) & 0.9856 (0.1427) \\
		& IPW & -0.0563 (0.1496) & 0.1644 (0.1506) & 0.3140 (0.1547) & 0.3662 (0.1671) & 0.3498 (0.1666) \\ [0.8em]
		\multirow{3}{2em}{-1} & EM & 0.0519 (0.1392) & 0.2600 (0.1381) & 0.4426 (0.1388) & 0.5801 (0.1372) & 0.6728 (0.1365) \\
		& stoEM & 0.0531 (0.1371) & 0.2566 (0.1379) & 0.4397 (0.1375) & 0.5835 (0.1365) & 0.6730 (0.1363) \\
		& IPW & -0.0563 (0.1496) & 0.1068 (0.1530) & 0.1991 (0.1591) & 0.2324 (0.1634) & 0.1995 (0.1677) \\ [0.8em]
		\multirow{3}{2em}{-0.5} & EM & 0.0583 (0.1337) & 0.1619 (0.1336) & 0.2516 (0.1335) & 0.3188 (0.1333) & 0.3641 (0.1330) \\
		& stoEM & 0.0567 (0.1334) & 0.1580 (0.1336) & 0.2528 (0.1332) & 0.3202 (0.1331) & 0.3636 (0.1330) \\
		& IPW & -0.0563 (0.1496) & 0.0196 (0.1548) & 0.0820 (0.1626) & 0.0844 (0.1648) & 0.0685 (0.1687) \\ [0.8em]
		\multirow{3}{2em}{0} & EM & 0.0605 (0.1319) & 0.0605 (0.1319) & 0.0605 (0.1319) & 0.0605 (0.1319) & 0.0605 (0.1319) \\
		& stoEM & 0.0605 (0.1318) & 0.0605 (0.1318) & 0.0605 (0.1318) & 0.0605 (0.1318) & 0.0605 (0.1318) \\
		& IPW & -0.0563 (0.1496) & -0.0703 (0.1540) & -0.0751 (0.1609) & -0.0792 (0.1676) & -0.0744 (0.1719) \\ [0.8em]
		\multirow{3}{2em}{0.5} & EM & 0.0591 (0.1336) & -0.0432 (0.1335) & -0.1306 (0.1333) & -0.1957 (0.1332) & -0.2396 (0.1332) \\
		& stoEM & 0.0561 (0.1331) & -0.0447 (0.1334) & -0.1338 (0.1332) & -0.1964 (0.1336) & -0.2407 (0.1332) \\
		& IPW & -0.0563 (0.1496) & -0.1584 (0.1556) & -0.2141 (0.1625) & -0.2314 (0.1772) & -0.2357 (0.1711) \\ [0.8em]
		\multirow{3}{2em}{1} & EM & 0.0538 (0.1382) & -0.1489 (0.1385) & -0.3215 (0.1377) & -0.4504 (0.1373) & -0.5375 (0.1375) \\
		& stoEM & 0.0584 (0.1376) & -0.1464 (0.1365) & -0.3160 (0.1371) & -0.4584 (0.1370) & -0.5373 (0.1364) \\
		& IPW & -0.0563 (0.1496) & -0.2453 (0.1529) & -0.3511 (0.1625) & -0.3908 (0.1612) & -0.3827 (0.1678) \\ [0.8em]
		\multirow{3}{2em}{1.5} & EM & 0.0432 (0.1453) & -0.2542 (0.1444) & -0.5084 (0.1439) & -0.7011 (0.1448) & -0.8322 (0.1430) \\
		& stoEM & 0.0322 (0.1441) & -0.2530 (0.1421) & -0.5164 (0.1417) & -0.6963 (0.1414) & -0.8396 (0.1421) \\
		& IPW & -0.0563 (0.1496) & -0.3192 (0.1530) & -0.4691 (0.1584) & -0.5321 (0.1607) & -0.5289 (0.1652) \\ [0.8em]
		\multirow{3}{2em}{2} & EM & 0.0268 (0.1527) & -0.3520 (0.1519) & -0.6806 (0.1535) & -0.9394 (0.1527) & -1.1181 (0.1508) \\
		& stoEM & 0.0280 (0.1523) & -0.3612 (0.1481) & -0.6909 (0.1454) & -0.9496 (0.1481) & -1.1195 (0.1466) \\
		& IPW & -0.0563 (0.1496) & -0.3878 (0.1508) & -0.5729 (0.1532) & -0.6322 (0.1574) & -0.6322 (0.1601) \\
\hline
\hline 
\end{tabular}
\end{center}
\label{tab:CDout1_neg2}
\end{table}
\end{landscape}

\begin{landscape}
\begin{table}[!thpb]
\small
\caption{Sensitivity analysis results on the IBD for CD patients data for outcome clinical remission with $\zeta_2 = 0$}
\begin{center}
\begin{tabular}{clccccccc}
\hline
\hline 
                  $\zeta_1$ & method & $\zeta^z = 0$   & $\zeta^z = 0.5$ & $\zeta^z = 1$ & $\zeta^z = 1.5$  & $\zeta^z = 2$    \\
\hline 
		\multirow{3}{2em}{-2} & EM & 0.0263 (0.1538) & 0.4227 (0.1544) & 0.7952 (0.1568) & 1.0920 (0.1551) & 1.2941 (0.1513) \\
		& stoEM & 0.0363 (0.1506) & 0.4126 (0.1491) & 0.7955 (0.1508) & 1.1013 (0.1487) & 1.2955 (0.1488) \\
		& IPW & -0.0563 (0.1496) & 0.2300 (0.1507) & 0.4010 (0.1537) & 0.4628 (0.1579) & 0.4488 (0.1638) \\ [0.8em]
		\multirow{3}{2em}{-1.5} & EM & 0.0427 (0.1469) & 0.3492 (0.1451) & 0.6257 (0.1462) & 0.8388 (0.1437) & 0.9833 (0.1429) \\
		& stoEM & 0.0406 (0.1420) & 0.3494 (0.1419) & 0.6260 (0.1422) & 0.8478 (0.1427) & 0.9840 (0.1426) \\
		& IPW & -0.0563 (0.1496) & 0.1641 (0.1507) & 0.3087 (0.1543) & 0.3622 (0.1695) & 0.3483 (0.1661) \\ [0.8em]
		\multirow{3}{2em}{-1} & EM & 0.0533 (0.1387) & 0.2591 (0.1381) & 0.4404 (0.1385) & 0.5781 (0.1372) & 0.6714 (0.1365) \\
		& stoEM & 0.0534 (0.1368) & 0.2558 (0.1380) & 0.4382 (0.1371) & 0.5801 (0.1366) & 0.6717 (0.1360) \\
		& IPW & -0.0563 (0.1496) & 0.1078 (0.1526) & 0.1956 (0.1602) & 0.2241 (0.1630) & 0.1965 (0.1678) \\ [0.8em]
		\multirow{3}{2em}{-0.5} & EM & 0.0588 (0.1336) & 0.1615 (0.1335) & 0.2507 (0.1333) & 0.3180 (0.1332) & 0.3636 (0.1330) \\
		& stoEM & 0.0581 (0.1332) & 0.1574 (0.1334) & 0.2536 (0.1332) & 0.3194 (0.1332) & 0.3627 (0.1329) \\
		& IPW & -0.0563 (0.1496) & 0.0210 (0.1554) & 0.0778 (0.1623) & 0.0802 (0.1643) & 0.0631 (0.1679) \\ [0.8em]
		\multirow{3}{2em}{0} & EM & 0.0605 (0.1319) & 0.0605 (0.1319) & 0.0605 (0.1319) & 0.0605 (0.1319) & 0.0605 (0.1319) \\
		& stoEM & 0.0605 (0.1318) & 0.0605 (0.1318) & 0.0605 (0.1318) & 0.0605 (0.1318) & 0.0605 (0.1318) \\
		& IPW & -0.0563 (0.1496) & -0.0703 (0.1545) & -0.0782 (0.1612) & -0.0826 (0.1692) & -0.0807 (0.1718) \\ [0.8em]
		\multirow{3}{2em}{0.5} & EM & 0.0588 (0.1336) & -0.0429 (0.1335) & -0.1301 (0.1333) & -0.1952 (0.1333) & -0.2393 (0.1332) \\
		& stoEM & 0.0561 (0.1330) & -0.0425 (0.1334) & -0.1327 (0.1334) & -0.1966 (0.1336) & -0.2400 (0.1332) \\
		& IPW & -0.0563 (0.1496) & -0.1597 (0.1559) & -0.2179 (0.1625) & -0.2411 (0.1769) & -0.2436 (0.1709) \\ [0.8em]
		\multirow{3}{2em}{1} & EM & 0.0533 (0.1383) & -0.1485 (0.1386) & -0.3207 (0.1378) & -0.4498 (0.1375) & -0.5371 (0.1374) \\
		& stoEM & 0.0576 (0.1377) & -0.1437 (0.1364) & -0.3150 (0.1370) & -0.4572 (0.1369) & -0.5379 (0.1363) \\
		& IPW & -0.0563 (0.1496) & -0.2455 (0.1530) & -0.3554 (0.1620) & -0.3976 (0.1618) & -0.3887 (0.1697) \\ [0.8em]
		\multirow{3}{2em}{1.5} & EM & 0.0427 (0.1453) & -0.2536 (0.1442) & -0.5075 (0.1440) & -0.7005 (0.1448) & -0.8319 (0.1430) \\
		& stoEM & 0.0301 (0.1447) & -0.2555 (0.1420) & -0.5189 (0.1418) & -0.6974 (0.1418) & -0.8379 (0.1414) \\
		& IPW & -0.0563 (0.1496) & -0.3192 (0.1534) & -0.4751 (0.1599) & -0.5378 (0.1609) & -0.5351 (0.1646) \\ [0.8em]
		\multirow{3}{2em}{2} & EM & 0.0263 (0.1527) & -0.3514 (0.1534) & -0.6800 (0.1540) & -0.9390 (0.1528) & -1.1179 (0.1508) \\
		& stoEM & 0.0286 (0.1512) & -0.3634 (0.1482) & -0.6890 (0.1467) & -0.9512 (0.1484) & -1.1178 (0.1472) \\
		& IPW & -0.0563 (0.1496) & -0.3878 (0.1509) & -0.5791 (0.1541) & -0.6408 (0.1574) & -0.6410 (0.1597) \\
\hline
\hline 
\end{tabular}
\end{center}
\label{tab:CDout1_0}
\end{table}
\end{landscape}

\begin{landscape}
\begin{table}[!thpb]
\small
\caption{Sensitivity analysis results on the IBD for CD patients data for outcome clinical remission with $\zeta_2 = 2$}
\begin{center}
\begin{tabular}{clccccccc}
\hline
\hline 
                  $\zeta_1$ & method  & $\zeta^z = 0$   & $\zeta^z = 0.5$ & $\zeta^z = 1$ & $\zeta^z = 1.5$  & $\zeta^z = 2$    \\
\hline  
		\multirow{3}{2em}{-2} & EM & 0.0268 (0.1533) & 0.4254 (0.1545) & 0.8006 (0.1568) & 1.0991 (0.1547) & 1.3010 (0.1509) \\
		& stoEM & 0.0349 (0.1488) & 0.4152 (0.1487) & 0.8002 (0.1506) & 1.1081 (0.1482) & 1.3077 (0.1490) \\
		& IPW & -0.0563 (0.1496) & 0.2313 (0.1508) & 0.3976 (0.1543) & 0.4547 (0.1568) & 0.4415 (0.1637) \\ [0.8em]
		\multirow{3}{2em}{-1.5} & EM & 0.0432 (0.1466) & 0.3516 (0.1457) & 0.6304 (0.1463) & 0.8450 (0.1435) & 0.9896 (0.1427) \\
		& stoEM & 0.0423 (0.1416) & 0.3551 (0.1425) & 0.6304 (0.1425) & 0.8537 (0.1426) & 0.9907 (0.1428) \\
		& IPW & -0.0563 (0.1496) & 0.1681 (0.1512) & 0.3073 (0.1551) & 0.3536 (0.1686) & 0.3372 (0.1649) \\ [0.8em]
		\multirow{3}{2em}{-1} & EM & 0.0538 (0.1387) & 0.2610 (0.1380) & 0.4440 (0.1386) & 0.5828 (0.1371) & 0.6764 (0.1365) \\
		& stoEM & 0.0537 (0.1366) & 0.2574 (0.1380) & 0.4411 (0.1373) & 0.5855 (0.1369) & 0.6780 (0.1358) \\
		& IPW & -0.0563 (0.1496) & 0.1106 (0.1534) & 0.1943 (0.1596) & 0.2151 (0.1626) & 0.1871 (0.1662) \\ [0.8em]
		\multirow{3}{2em}{-0.5} & EM & 0.0591 (0.1335) & 0.1626 (0.1334) & 0.2528 (0.1333) & 0.3207 (0.1332) & 0.3664 (0.1330) \\
		& stoEM & 0.0582 (0.1331) & 0.1595 (0.1332) & 0.2553 (0.1331) & 0.3225 (0.1330) & 0.3657 (0.1328) \\
		& IPW & -0.0563 (0.1496) & 0.0202 (0.1559) & 0.0758 (0.1618) & 0.0735 (0.1647) & 0.0567 (0.1664) \\ [0.8em]
		\multirow{3}{2em}{0} & EM & 0.0605 (0.1319) & 0.0605 (0.1319) & 0.0605 (0.1319) & 0.0605 (0.1319) & 0.0605 (0.1319) \\
		& stoEM & 0.0605 (0.1318) & 0.0605 (0.1318) & 0.0605 (0.1318) & 0.0605 (0.1318) & 0.0605 (0.1318) \\
		& IPW & -0.0563 (0.1496) & -0.0688 (0.1546) & -0.0797 (0.1611) & -0.0907 (0.1691) & -0.0893 (0.1717) \\ [0.8em]
		\multirow{3}{2em}{0.5} & EM & 0.0583 (0.1337) & -0.0446 (0.1337) & -0.1328 (0.1333) & -0.1986 (0.1333) & -0.2427 (0.1332) \\
		& stoEM & 0.0560 (0.1332) & -0.0435 (0.1333) & -0.1354 (0.1334) & -0.2005 (0.1335) & -0.2437 (0.1332) \\
		& IPW & -0.0563 (0.1496) & -0.1567 (0.1567) & -0.2186 (0.1616) & -0.2461 (0.1781) & -0.2511 (0.1700) \\ [0.8em]
		\multirow{3}{2em}{1} & EM & 0.0519 (0.1384) & -0.1523 (0.1387) & -0.3270 (0.1381) & -0.4571 (0.1373) & -0.5440 (0.1375) \\
		& stoEM & 0.0566 (0.1380) & -0.1491 (0.1363) & -0.3217 (0.1368) & -0.4649 (0.1372) & -0.5443 (0.1361) \\
		& IPW & -0.0563 (0.1496) & -0.2458 (0.1544) & -0.3561 (0.1625) & -0.4040 (0.1614) & -0.3972 (0.1708) \\ [0.8em]
		\multirow{3}{2em}{1.5} & EM & 0.0403 (0.1464) & -0.2600 (0.1451) & -0.5176 (0.1445) & -0.7115 (0.1441) & -0.8414 (0.1429) \\
		& stoEM & 0.0296 (0.1454) & -0.2664 (0.1410) & -0.5270 (0.1423) & -0.7087 (0.1425) & -0.8466 (0.1408) \\
		& IPW & -0.0563 (0.1496) & -0.3153 (0.1538) & -0.4771 (0.1599) & -0.5403 (0.1616) & -0.5395 (0.1634) \\ [0.8em]
		\multirow{3}{2em}{2} & EM & 0.0229 (0.1531) & -0.3591 (0.1548) & -0.6929 (0.1540) & -0.9521 (0.1528) & -1.1279 (0.1507) \\
		& stoEM & 0.0294 (0.1544) & -0.3707 (0.1471) & -0.6994 (0.1474) & -0.9558 (0.1466) & -1.1310 (0.1484) \\
		& IPW & -0.0563 (0.1496) & -0.3869 (0.1511) & -0.5796 (0.1543) & -0.6431 (0.1585) & -0.6405 (0.1590) \\
\hline
\hline 
\end{tabular}
\end{center}
\label{tab:CDout1_pos2}
\end{table}
\end{landscape}

\end{document}